\documentclass[letterpaper,11pt]{article}
\pdfoutput=1
\usepackage{jheppub}
\pdfoutput=1
\usepackage{amsmath}
\usepackage{amsfonts}
\usepackage{amssymb}
\usepackage{color}

\def\be{\begin{equation}}
\def\ee{\end{equation}}
\def\bea{\begin{eqnarray}}
\def\eea{\end{eqnarray}}

\def\ordr{\mathcal{O}}
\def\lag{\mathcal{L}}
\def\vep{\varepsilon}
\def\<{\langle}
\def\>{\rangle}
\def\non{\nonumber}
\def\TD{\text{TD}}
\def\gal{\text{Gal}}
\def\sgal{\text{sGal}}
\def\mt{\mathcal{T}}

\def\J{\mathcal{J}}
\def\dbi{\text{DBI}}
\def\tr{\text{Tr}}
\def\gs{\Sigma}
\def\gst{\tilde{\Sigma}}
\def\ta{{\tilde{a}}}
\def\tb{{\tilde{b}}}
\def\tc{{\tilde{c}}}
\def\nlsm{\text{NLSM}}

\def\nf{\mathbf{f}}

\def\pf {\mathrm{Pf}}
\def\A {\textsf{A}}
\def\mphi{\underline{\phi}}
\def\mpi{\underline{\pi}}
\def\mgs{\underline{\gs}}
\def\mgst{\underline{\gst}}
\def\mPhi{\underline{\Phi}}
\def\nV{\mathcal{V}}
\def\mPsi{\underline{\Psi}}

\newcommand*\DAl{\mathop{}\!\mathbin\Box}

\title{The Infrared Structure of Exceptional Scalar Theories}

\author{Zhewei Yin}
\affiliation{Department of Physics and Astronomy, Northwestern University, Evanston, IL 60208, USA}
\emailAdd{zheweiyin2015@u.northwestern.edu}

\abstract{Exceptional theories are a group of one-parameter scalar field theories with (enhanced) vanishing soft limits in the S-matrix elements. They include the nonlinear sigma model (NLSM), Dirac-Born-Infeld scalars and the special Galileon theory. The soft behavior results from the shift symmetry underlying these theories, which leads to Ward identities generating subleading single soft theorems as well as novel Berends-Giele recursion relations. Such an approach was first applied to NLSM in Refs. \cite{Low:2017mlh,Low:2018acv}, and here we use it to systematically study other exceptional scalar field theories. In particular, using the subleading single soft theorem for the special Galileon we identify the Feynman vertices of the corresponding extended theory, which was first discovered using the Cachazo-He-Yuan representation of scattering amplitudes.  Furthermore, we present a Lagrangian for the extended theory of the special Galileon, which has a rich particle content involving biadjoint scalars, Nambu-Goldstone bosons and Galileons, as well as additional flavor structure.}

\begin{document}

\maketitle

\flushbottom

\section{Introduction}
\label{sect:intro}

There is a long history of research on the infrared structure of scattering amplitudes \cite{Low:1954kd,GellMann:1954kc,Low:1958sn,Burnett:1967km,Weinberg:1964ew,Weinberg:1965nx,Gross:1968in,Jackiw:1968zza,Adler:1964um,Treiman:1986ep}. The results are mainly expressed in terms of ``soft theorems'', which describe the asymptotic behavior of scattering amplitudes when the momentum of one or several external legs approaches zero. Prominent examples include soft theorems for gauge theory and gravity \cite{Low:1958sn,Weinberg:1965nx}, as well as the Adler's zero condition for Nambu-Goldstone (NG) bosons \cite{Adler:1964um}.

Recently there is renewed interest in studying the infrared structure of scattering amplitudes \cite{ArkaniHamed:2008gz,Strominger:2013lka,Strominger:2013jfa,He:2014cra}. New soft theorems for a higher order of soft momenta in gauge theory and gravity were discovered \cite{Cachazo:2014fwa,Casali:2014xpa}, which sparked many new analyses (see, e.g. Ref. \cite{Strominger:2017zoo} and various references therein). The soft theorems of gauge theory and gravity have long been associated with gauge invariance in the theory, and it is recently realized that more specifically, they are the manifestation of large gauge transformation that does not fall off at infinity \cite{Strominger:2017zoo,Hamada:2018vrw}. Much of the recent work focuses on analyzing the interplay between this kind of  asymptotic symmetry and the soft behavior of scattering amplitudes.

The story is slightly different for scalar effective field theories (EFTs). It was recognized early on that the Adler's zero for NG bosons (NGBs) is a consequence of the spontaneously broken symmetry \cite{Adler:1964um,Dashen:1969ez,Weinberg:1966gjf,Weinberg:1970bs}, which manifests as shift symmetry for the NGBs. On the other hand, there were also early attempts which used the soft behavior of the amplitudes as the defining property of the scalars, and tried to construct the Lagrangian for the whole theory and recover the broken symmetries \cite{Susskind:1970gf,Ellis:1970nn}. This idea has been fully realized in Refs. \cite{Cheung:2015ota,Cheung:2016drk}, where on-shell recursion relations constrained by soft behavior of scattering amplitudes were used to build up all tree-level S-matrix elements. Although different recursion relations for nonlinear sigma model (NLSM) were already proposed in Refs. \cite{Kampf:2012fn,Kampf:2013vha}, the soft behavior of scattering amplitudes becomes explicit in recursion relations given by Ref. \cite{Cheung:2015ota}.

Moreover, apart from the well-studied NLSM, other theories with non-trivial soft behavior have been discovered, including the Dirac-Born-Infeld (DBI) scalars and the special Galileon (sGal) theory \cite{Cheung:2014dqa,Hinterbichler:2015pqa}, which is a special case of the Galileon theory (Gal) \cite{Nicolis:2008in}. The on-shell amplitudes of these theories possess ``enhanced'' Adler's zero. To be concrete, when we take an external momentum $q$ of an on-shell amplitude to be soft, we replace $q^\mu$ with $\tau q^\mu$ and take the limit of $\tau \to 0$. Then enhanced Adler's zero means that the leading term in the single soft limit of the scattering amplitude is at $\ordr (\tau^s)$ with $s>1$. As the soft degree $s$ in these theories are also larger than what the naive power counting of derivatives in the Lagrangian would give, they were called ``exceptional theories'' in Refs. \cite{Cheung:2014dqa,Cheung:2016drk}. As for special Galileon, the shift symmetry that protects the soft behavior was discovered \cite{Hinterbichler:2015pqa} after the enhanced soft behavior was identified, and the relation between the internal symmetry and enhanced soft behavior was further clarified in Refs. \cite{Cheung:2016drk,Bogers:2018zeg}.

A different line of research utilized the newly developed tool of Cachazo-He-Yuan (CHY) representation of scattering amplitudes \cite{Cachazo:2013gna,Cachazo:2013hca,Cachazo:2013iea,Cachazo:2014xea}. Exceptional scalar EFTs of one parameters, including NLSM, DBI and special Galileon, arise naturally in the CHY formalism for different reasons \cite{Cachazo:2014xea}. Apart from the Adler's zero in the leading order  soft limit as well as new double soft limits \cite{Cachazo:2015ksa}, subleading single soft results for NLSM and special Galileon were also derived \cite{Cachazo:2016njl}, where new ``extended theories'' of different kinds of scalars interacting with each other emerge. These mixed theories had not been studied before, and only the CHY representation of the amplitudes in these theories were given. A recursion relation for NLSM different from Ref. \cite{Cheung:2015ota} was also constructed, where the scattering amplitudes of the extended theory and NLSM are intertwined.

The soft limit of scalar EFTs has also been studied using traditional quantum field theory methods \cite{Du:2015esa,Du:2016njc,Low:2015ogb}. In particular, in Refs. \cite{Low:2014nga,Low:2014oga,Low:2017mlh,Low:2018acv} the Adler's zero in NLSM is again imposed as a defining property for the scalar EFT. However, instead of on-shell recursion relations, the Lagrangian and the complete form of the shift symmetry are directly generated, and a Ward identity for correlation functions naturally arises. The Ward identity gives rise to new recursion relations for semi-on-shell amplitudes, so that not only the Adler's zero becomes manifest,  the subleading soft limit for NLSM can also be derived straightforwardly. The corresponding amplitudes and Feynman rules of the extended theory in Ref. \cite{Cachazo:2016njl} were easily identified, which shed new light on the origin and property of the interaction between different scalars in the extended theory.

In this work we adopt a similar approach. We consider exceptional theories with enhanced soft behaviors, including DBI and special Galileon. As the shift symmetry corresponding to the enhanced Adler's zero in these theories have already been identified, we directly use them to generate Ward identities. Apart from showing explicitly that the leading soft behavior in these theories is directly connected to the Ward identity of the shift symmetry, we are also able to provide new recursion relations and subleading single soft theorems of these theories. For the case of special Galileon, we also identify the amplitudes of the extended theory discovered in Ref. \cite{Cachazo:2016njl} in the subleading single soft limit, and we provide Feynman rules for the extended theory. We also  identify the Lagrangian that generates Feynman rules of the extended theory, which leads to many open questions.

The paper is organized as follows. We present the Ward identity and subleading single soft limit of scalar EFTs in Section \ref{sec:WI}. We start with a brief review of results in Ref. \cite{Low:2017mlh,Low:2018acv} on NLSM, then approach the cases of DBI, ordinary Galileon and special Galileon, providing new recursion relations as well as subleading single soft theorems. In Section \ref{sec:ext} we study the amplitudes of the extended theories, by first discussing what we learn from the NLSM case, and then identifying the amplitudes and Feynman rules of the extended theory in the soft limit of special Galileon.  We then proceed to propose the Lagrangian of the extended theory. We conclude and discuss the outlook in Section \ref{sec:conclusion}. We also include four appendices: Appendix \ref{app:gall} presents properties of the Galileon Lagrangian, which are used to derive the results of (special) Galileon theory, while Appendix \ref{app:vanlsm} discusses details of NLSM including flavor-ordering and parameterization. A necessary review of the CHY formulae is presented in Appendix \ref{app:chyr}, and detailed derivation of Feynman vertices in the extension of special Galileon theory is presented in Appendix \ref{app:sgaled}.

\section{The Ward identity and the subleading single soft limit}
\label{sec:WI}

It has long been recognized that the Adler's zero in scattering amplitudes of NGBs can be derived from current conservation corresponding to the nonlinear shift symmetry. The textbook example \cite{Weinberg:1996kr} is that given a scalar $\pi(x)$ in a theory with the shift symmetry, namely the invariance under the transformation
\bea
\pi (x) \to \pi (x) + \vep,
\eea
where $\vep $ is a constant, one can associate a Noether current $\J^\mu (x)$ with the symmetry. The current contains a one-particle pole:
\bea
\< \Omega | \J^\mu (x) | \pi (p) \> = i f p^\mu e^{-i p\cdot x},
\eea
where $f $ is a coupling constant similar to the pion decay constant in the QCD chiral Lagrangian. Current conservation corresponding to the shift symmetry leads to the Ward identity:
\bea
\partial_\mu \<\ \widehat{f}\ | \J^\mu (x) |\ \widehat{i}\ \> = 0 \label{eq:gwi0}
\eea
for general on-shell initial and final states $\widehat{i}$ and $\widehat{f}$. After Fourier transforming $\J^\mu (x)$ to $\tilde{\J}^\mu (p)$ and singling out the one-particle contribution, the equation above becomes
\bea
\<\ \widehat{f}\ + \pi (p)| \ \widehat{i}\ \> = p_\mu R^\mu (p),\label{eq:gwi}
\eea
where the remainder function $R^\mu (p)$ is the matrix element of $\tilde{\J}^\mu (p)$ with the one particle pole removed, thus $\lim_{p \to 0} p^2 R^\mu (p) = 0$. To prove the Adler's zero condition, we need a stronger regularity condition for $R^\mu (p) $, namely $\lim_{p \to 0} p^\nu R^\mu (p) = 0$, which can be ensured if there is no cubic vertex in the theory. Then we have
\bea
\lim_{p \to 0} \<\ \widehat{f} \ + \pi (p)| \ \widehat{i}\ \> = \lim_{p \to 0} p_\mu R^\mu (p) = 0,
\eea
which is the Adler's zero for on-shell amplitudes.

The discussion above can be generalized to scalars with flavor, e.g. in NLSM, or enhanced soft limit, e.g. in DBI, ordinary Galileon and special Galileon. The case for general enhanced Adler's zero has been discussed in  \cite{Cheung:2016drk}. It is also clear from our discussion that  to get the next leading order soft theorem, one simply needs to calculate the remainder function $R_\mu (p) $ to the lowest order in $p$. In the following sections we will realize this idea for NLSM, DBI, ordinary Galileon and special Galileon.

\subsection{Nonlinear Sigma Model}
\label{sec:nlsm}

In the modern formulation of NLSM by Coleman, Callan, Wess and Zumino (CCWZ) \cite{Coleman:1969sm,Callan:1969sn}, the scalars are NGBs parameterizing a coset space $G/H$. The  spontaneously broken group $G$ contains the unbroken subgroup $H$, so that the generators of $G$ are divided into two classes: the ``unbroken generators'' $T^i$ of the subgroup $H$, and the ``broken generators'' $X^a$ associated with the coset $G/H$. For the global internal symmetry $G$, each generator $X^a$ corresponds to an NGB $\pi^a$, and the knowledge of both the unbroken and broken generators is mandatory to construct the NLSM Lagrangian. In other words, the CCWZ formalism relies on the coset space $G/H$, so that it is not enough to  know the unbroken symmetry in the infrared to construct the Lagrangian: for a different $G$, the Lagrangian is seemingly different even if $H$ is the same.

As we will see shortly, to compute the function $R^\mu$ in Eq. (\ref{eq:gwi}) requires a complete knowledge of the shift symmetry. For NLSM, the Lagrangian is invariant under the transformation \cite{Low:2014nga}
\bea
\pi^a(x) \to \pi^a(x) + F_1^{ab}  \vep^b,\label{eq:nlsmsg}
\eea
where $\vep^a$ are arbitrary constants, and $F_1^{ab} = \delta^{ab} + \ordr (1/f^2)$ is a function containing the field $\pi^a(x)$. Early on NLSM is mainly studied using current algebraic techniques \cite{Dashen:1969ez}, which makes it hard to calculate $F_1$ beyond the leading order in $1/f^2$. Another obstacle is related to the fact that in the CCWZ formalism, the theory seems to depend on the broken group $G$. For example, in Ref. \cite{Ellis:1970nn} the relation between $F_1$ and the form of Lagrangian is established for the simple case of $H= SU(2)$, but the relation seems different for three different possible $G$, namely $G=SU(2 )\times SU(2)$, $E(3)$ or $SO(3,1)$.

It is realized recently in Refs. \cite{Low:2014nga,Low:2014oga} that in contrast to what the traditional CCWZ formalism may imply, the nonlinearity in NLSM can be completely fixed by the infrared information, including the unbroken group $H$ and the soft behavior of the scattering amplitudes, i.e. Adler's zero condition. Choosing a basis where the generators of $H$ are purely imaginary and antisymmetric, the Lagrangian at the two derivative level can be completely fixed to be\footnote{A ``closure condition'' \cite{Low:2014nga} of $T^i$ needs to be imposed, which is universal and can be completely fixed by infrared information, and amounts to requiring $H$ to be able to be embedded into a symmetric coset.}
\bea
\lag_{\text{NLSM}}^{(2)} = \frac{1}{2} \partial^\mu \pi^a \partial_\mu \pi^b \left[ F_2 (\mt)^2 \right]^{ab},\label{eq:nlsmlag}
\eea
where
\bea
F_2 (\mt) = \frac{\sin \sqrt{\mt} }{\sqrt{\mt} }, \qquad (\mt)_{ab} = \frac{1}{f ^2} T^i_{ac} T^i_{db} \pi^c \pi^d.
\eea
 From the form of $F_2$ we know that the Lagrangian in Eq. (\ref{eq:nlsmlag}) does not depend on broken generators $X$, therefore its nonlinearity is independent of the coset. The only dependence on the coset $G/H$ is contained in the normalization of the NGB field $\pi^a$, i.e. the coupling constant $f $. For example, the three cases in Ref. \cite{Ellis:1970nn} for $G = SU(2) \times SU(2)$, $E(3)$ and $SO(3,1)$ correspond to $1/f^2 >0$, $=0$ and $<0$, respectively.

By the same argument, the nonlinear shift in Eq. (\ref{eq:nlsmsg}) should have a form independent of $G/H$ as well. The function $F_1$ is found to be \cite{Low:2017mlh,Low:2018acv}
\bea
F_1 (\mt) = \sqrt{\mt} \cot \sqrt{\mt}.
\eea
Now we are ready to use the Lagrangian in Eq. (\ref{eq:nlsmlag})  to calculate the Ward identity associated with the shift symmetry. First, we promote the global shift to a local one: $\vep^a \to \vep^a (x)$. As the Lagrangian is invariant under the shift symmetry, all the terms in the shifted Lagrangian $\lag'$ that are linear in $\vep$ will vanish, so that to the leading order in $\vep^a$,
\begin{align}
 \mathcal{L} [\pi'] =\mathcal{L} [\pi] + \frac{\delta \mathcal{L}}{\delta \pi^a} \delta \pi^a + \frac{\delta \mathcal{L}}{\delta(\partial_\mu  \pi^a)} \delta(\partial_\mu \pi^a) = \mathcal{L} [\pi] +  \partial^\mu \pi^a  \left[ F_3( \mt) \right]^{ab} \partial_\mu \vep^b,
\end{align}
where
\bea
F_3( \mt) = \frac{\sin \sqrt{\mt} \cos \sqrt{\mt}}{\sqrt{\mt}}.
\eea
Classically, the action does not change when we have a small variation of the field due to the equation of motion, thus the change in action is
\bea
\delta S = \int d x^4\ \delta \lag = \int d x^4\  \partial^\mu \pi^a  \left[ F_3( \mt) \right]^{ab} \partial_\mu \vep^b = 0,
\eea
which leads to conservation of current: $\partial^\mu \J^a_\mu = 0$, with
\bea
\J_\mu^a =   \left[F_3(\mt) \right]^{ab}\partial_\mu \pi^b = \partial_\mu \pi^a + \sum_{k=1}^\infty \frac{(-4)^k}{(2k+1)!} (\mt^k)_{ab} \partial_\mu \pi^b.\label{eq:nlsmc}
\eea
Suppose the conservation of current survives quantization,\footnote{It is known that the normal Ward identities can be modified by quantum anomalies \cite{Adler:1969gk}.} we arrive at the Ward identity
\begin{align}
&i \partial^\mu \< \Omega| \J^a_\mu (x) \prod_{i=1}^n \pi^{a_i} (x_i) |\Omega \>\nonumber \\
&=  \sum_{r=1}^n \<\Omega| \pi^{a_1} (x_1) \cdots \left[F_1(\mt)\right]_{a_r a}(x_r) \delta^{(4)}(x-x_r) \cdots \pi^{a_n} (x_n) |\Omega \> \label{eqnlsmwi}.
\end{align}
Both of the functions $F_1$ and $F_3$ contain information of the shift, making it clear that we need the complete form of the shift to derive the Ward identity. We see explicitly in Eq. (\ref{eq:nlsmc}) that the current $\J_\mu^a$ contains a one-particle pole, as well as terms of higher orders in $1/f^2 $ that will contribute to the remainder function.

To turn the Ward identity into a relation for on-shell amplitudes, we perform the Lehmann-Symanzik-Zimmermann (LSZ) reduction by integrating both sides of Eq. (\ref{eqnlsmwi}) using
\bea
\left(\frac{i}{\sqrt{Z}}\right)^{n} \int d^4 x\, e^{-i q \cdot x}  \prod_{i=1}^n \lim_{p_i^2 \to 0} \int d^4 x_i\, e^{-i p_i \cdot x_i}\, \DAl_i ,
\eea
where the d'Alembertian $\DAl_i$ acting on coordinate $x_i$ will eliminate the one particle pole $1/p_i^2$ on the left hand side (LHS) of Eq. (\ref{eqnlsmwi}), and each term on the right hand side (RHS) of Eq. (\ref{eqnlsmwi}) readily vanishes as the one particle pole of $\pi^{a_r} (x_r)$ is missing. The Fourier transformation on $x$ imposes the momentum conservation condition $q = -\sum_{i=1}^n p_i$. The Ward identity then becomes
\bea
&&q^2 J^{a_1 \cdots a_n, a} (p_1, \cdots ,p_n) \non\\
 &=& \sum_{k=1}^{\infty} \frac{(-4)^k}{(2k+1)!}   \<0| \int d^4 x e^{-i q \cdot x} \left[{\cal T}^k(x)\right]_{ab} iq \cdot \partial \pi^b(x)  | \pi^{a_1}(p_1) \cdots \pi^{a_n}(p_n)\> \ .\label{eq:nlsmqwi}
\eea
The LHS of Eq. (\ref{eq:nlsmqwi}) comes from the matrix element of the one particle pole in $\J_\mu^a$, where
\bea
J^{a_1 \cdots a_n, a} (p_1, \cdots ,p_n) = \langle 0| \pi^a (0) | \pi^{a_1}(p_1) \cdots \pi^{a_n}(p_n)\rangle
\eea
is the so-called ``semi-on-shell amplitude'', in which the legs with flavor indices $a_i$ and momenta $p_i$ are on-shell, with an additional off-shell leg of index $a$ and momentum $-\sum_{i=1}^n p_i = q$. The remaining part of  $\J_\mu^a$ without the one particle pole leads to the RHS of Eq. (\ref{eq:nlsmqwi}). Taking the on-shell limit, we arrive at
\bea
M^{a_1\cdots a_{n} a} (p_1, \cdots, p_n,  q) =  q \cdot R^{a_1 \cdots a_{n} a} (p_1, \cdots, p_n;  q),\label{eqnlsmssql}
\eea
with the remainder function
\bea
&& R_\mu^{a_1 \cdots a_{n} a} (p_1, \cdots, p_n;  q)\non\\
& = & \frac{1}{\sqrt{Z}} \sum_{k=1}^{\infty} \frac{-i(-4)^{k}}{(2k+1)!}  \<0| \int d^4x\, e^{-i q \cdot x} [{\cal T}^k(x)]_{a b}\   \partial_\mu \pi^b(x) | \pi^{a_1} \cdots \pi^{a_n}\> .\label{eq:nlsmR}
\eea
It is clear that Eq. (\ref{eqnlsmssql}) is a generalization of Eq. (\ref{eq:gwi}). The integrand in Eq. (\ref{eq:nlsmR}) can be seen as new $(2k+1)$-pt vertices with momentum insertion $q$, so that there is no danger of developing a pole when $q \to 0$, and the regularity condition $\lim_{q \to 0}q \cdot R  = 0$ holds. Therefore, Eq. (\ref{eqnlsmssql}) gives us Adler's zero condition, and taking the soft limit of $q$, the subleading single soft theorem is simply
\bea
M^{a_1\cdots a_{n} a} (p_1, \cdots, p_n,  \tau q) =  \tau q \cdot R^{a_1 \cdots a_{n} a} (p_1, \cdots, p_n;  0) + \ordr (\tau^2),\label{eq:nlsmslg}
\eea
which is valid at quantum level.\footnote{To be consistent, at loop level we also need to include terms in the Lagrangian with more than 2 derivatives, and the generalization of our method for such a  Lagrangian is straightforward.} Note that when $q=0$, the new vertices in Eq. (\ref{eq:nlsmR}) receive no momentum injection and become ``normal'' vertices, so that interpreting the leading term in Eq. (\ref{eq:nlsmslg}) as on-shell amplitudes becomes possible.

At tree level, the semi-on-shell amplitudes like the one in Eq. (\ref{eq:nlsmqwi}) are useful for constructing recursion relations, and they were first proposed by Berends and Giele to study Yang-Mills theories \cite{Berends:1987me}. Eq. (\ref{eq:nlsmqwi}) leads to a novel Berends-Giele type recursion relation \cite{Low:2017mlh,Low:2018acv}, where the RHS becomes semi-on-shell sub-amplitudes connected by the aforementioned new vertices. We will see in the following that such a Berends-Giele recursion relation is a common byproduct of the Ward identity, and in contrast to relations purely derived using Feynman rules, such as the one in Ref. \cite{Kampf:2013vha}, the new vertices are proportional to $s$ powers of the off-shell momentum $q$, making it very easy to identify the soft behavior of on-shell amplitudes. Here in NLSM we have $s=1$, while the Ward identities corresponding to enhanced shift symmetry in DBI, ordinary Galileon and special Galileon will generate vertices with $s>1$, enabling us to identify the enhanced Adler's zero.

\subsection{Dirac-Born-Infeld Scalars}
\label{sec:dbi}

The DBI action \cite{Leigh:1989jq} can be obtained by dimensional reduction from the Born-Infeld action \cite{Tseytlin:1999dj}, which describes a non-linear generalization of Maxwell theory, and can arise from string theory.
For a single flavor of scalar $\pi$, the DBI Lagrangian becomes
\bea
\lag_{\dbi} = -F^d \sqrt{1 - \frac{(\partial \pi)^2}{F^d}} + F^d.\label{eq:dbilag}
\eea
We will work in 4 dimensions so that $d  = 4$, although our methods can be easily extended to a general $d$. As there is explicitly a derivative acting on each of the $\pi$ in Eq. (\ref{eq:dbilag}), the Lagrangian is invariant under the constant shift $\pi \to \pi + \vep$. Applying the same method as discussed in Section \ref{sec:nlsm}, we can derive the Ward identity corresponding to the constant shift symmetry:
\bea
i \partial_\mu \< \J^\mu_{\dbi} (x) \prod_{i=1}^n \pi (x_i) \> = \sum_{r=1}^n \<\pi (x_1) \cdots \delta^{(4)}(x -x_r) \cdots \pi (x_n) \>.\label{eqcswi}
\eea
where the current is given by
\bea
\J^\mu_\dbi  = \frac{\partial^\mu \pi}{\sqrt{1 - (\partial \pi)^2/F^4}} = \partial^\mu \pi + \partial^\mu \pi \sum_{k=1}^\infty  \frac{(2k)!}{2^{2k} (k!)^2  } \left( \frac{\partial \pi}{F^2} \right)^2.\label{eq:dbicsj}
\eea
Again, the current contains a one-particle pole, as well as remainder terms with at least three fields. Then the remainder function,
\bea
R^\mu_{\dbi} (p_1, \cdots, p_n;  q) = -\frac{1}{\sqrt{Z}} \sum_{k=1}^{\infty} \frac{i(2k)!}{2^{2k} (k!)^2  }    \int d^4 x\,  \<0|  \left[  \frac{\partial \pi (x)}{F^{2}} \right]^{2k}  \partial^\mu \pi (x) | \pi (p_1) \cdots \pi (p_n)\>,\qquad
\eea
satisfies the regularity condition $\lim_{q\to 0} q \cdot R_{\dbi} (p_1, \cdots, p_n;  q)=0$. Therefore, taking the soft limit of $q$, the on-shell amplitude can be expressed as
\bea
M(p_1, \cdots, p_n, \tau q) &=&\tau q \cdot R_\dbi (p_1, \cdots, p_n;  0) + \ordr(\tau^2).\label{eqdbicssb}
\eea

It is clear that Adler's zero is manifest  in the Ward identity under the constant shift. However, DBI is an exceptional theory with the enhanced Adler's zero. The RHS of Eq. (\ref{eqdbicssb}) actually starts at $\ordr(\tau^2)$, but this behavior is not manifest in the Ward identity corresponding to the constant shift. This does not surprise us, as we already know that the enhanced Adler's zero is protected by the invariance of the theory under the enhanced shift  \cite{Cheung:2016drk}
\bea
\pi \to \pi +   \theta_\mu \left( x^\mu - F^{-d} \pi \partial^\mu \pi \right).\label{eq:dbis}
\eea
Unlike the case in NLSM, the Lagrangian is not invariant under the shift, but transforms as
\bea
 \lag_\dbi \to \lag_\dbi + \theta_\mu \partial^\mu D,\qquad D= \pi \sqrt{1- \frac{(\partial \pi)^2}{F^d}} .\label{eqeslv}
\eea
However, the action is still invariant under the enhanced shift. To calculate the current $j^{(\mu)}_\nu$ corresponding to the enhanced shift symmetry so that we have the current conservation condition $\partial^\nu j^{(\mu)}_\nu = 0$, we need to promote the constant $\theta_\mu$ in Eq. (\ref{eq:dbis}) to a function $\theta_\mu (x)$, and in the end arrive at
\bea
\delta S = \int d x^4 \ j^{(\mu)}_\nu \partial^\nu \theta_\mu = 0.
\eea
However, there is an alternative way to calculate the current: we make a general shift $\pi(x) \to \pi(x) + \vep(x)$ and calculate the variation of the action:
\bea
\delta S =\int dx^4 \ \J_\dbi^\mu \partial_\mu \vep = 0,
\eea
which leads to conservation of the current under the constant shift. As the function $\vep(x)$ is completely arbitrary, we can then replace $\vep(x)$ with the enhanced shift $\theta_\nu (x) \left( x^\mu - F^{-4} \pi \partial^\mu \pi \right)$, which leads to the relation between the two different currents \cite{Brauner:2014aha,Cheung:2016drk}:
\bea
\partial^\nu  j^{(\mu)}_\nu = \left( x^\mu - F^{-4} \pi \partial^\mu \pi \right) \partial_\nu \J_\dbi^\nu.\label{eq:dbijr}
\eea
Plugging in $\J_\dbi^\mu$ given in Eq. (\ref{eq:dbicsj}), we see that the RHS of Eq. (\ref{eq:dbijr}) can indeed be written as a total derivative:
\bea
\partial^\nu  j^{(\mu)}_\nu =  \partial_\nu \left[ (x^\mu -  F^{-4} \pi \partial^\mu \pi ) \J^\nu_\dbi - g^{\mu \nu} D \right],\label{eq:dbiej}
\eea
where the form of $D$ is given in Eq. (\ref{eqeslv}). We have the Ward identity corresponding to the enhanced shift symmetry:
\bea
i \partial^\nu \< j_\nu^{(\mu )} (x) \prod_{i=1}^n \pi (x_i) \> = \sum_{r=1}^n \<\pi (x_1) \cdots \left[ x^\mu - F^{-4} \pi (x) \partial^\mu \pi (x)  \right] \delta^{(4)}(x -x_r) \cdots \pi (x_n) \>.\label{eqeswip}
\eea

We should stress that we can derive the current directly without using the current relation in Eq. (\ref{eq:dbijr}), by promoting constant $\theta^\mu$ to $\theta^\mu (x)$ and including the change of the Lagrangian in Eq. (\ref{eqeslv}). These two methods are equivalent, while we will see that in the Galileon case it is much more convenient to use the current relation instead of calculating the current directly.

We can eliminate the $x^\mu$ term in Eq. (\ref{eq:dbiej}) so that we do not have derivatives of momentum when we express the Ward identity in terms of on-shell amplitudes. The trick is to combine Eq. (\ref{eqeswip}) with the Ward identity under the constant shift, Eq. (\ref{eqcswi}), which leads to
\bea
&&i \partial_\mu \partial_\nu\<{ j'}^{(\mu) \nu}(x) \prod_{i=1}^n \pi (x_i) \>\non\\
&=& \sum_{r=1}^n \<\pi (x_1) \cdots  \left\{  \delta^{(4)}(x -x_r)  + \partial_\mu \left[ F^{-4} \pi (x) \partial^\mu \pi (x) \delta^{(4)}(x -x_r) \right] \right\} \cdots \pi (x_n) \>,\label{eqdbifwi}
\eea
where we have contracted another total derivative with the free Lorentz index $\mu $ in Eq. (\ref{eq:dbiej}), and
\bea
\partial_\mu \partial_\nu { j'}^{(\mu) \nu} = \partial_\mu \partial_\nu [x^\mu \J_\dbi^\nu  - j^{(\mu) \nu} ]= \partial_\mu \partial_\nu [g^{\mu \nu} D  + F^{-4} \pi \partial^\mu \pi \J^\nu_\dbi  ].
\eea
There are 2 powers of total derivatives on the LHS of Eq. (\ref{eqdbifwi}). Normally a Ward identity contains one power of total derivative, as shown in Eq. (\ref{eq:gwi0}); because the enhanced shift given by Eq. (\ref{eq:dbis}) involves space-time coordinates, the parameter $\theta^\mu$ and consequently the current $j_\nu^{(\mu)}$ contains an additional Lorentz index, so that we can contract an additional total derivative in the Ward identity, which leads to the enhanced soft behavior. Indeed, by performing the LSZ reduction we can calculate the semi-on-shell amplitude of DBI as
\bea
q^2 J (p_1, \cdots ,p_n)  =- \sqrt{Z} q_\mu q_\nu r^{\mu \nu} (p_1, \cdots, p_n;  q),\label{eq:dbisaq}
\eea
with the remainder function $r^{\mu \nu} (p_1, \cdots, p_n;  q)$ given by
\bea
r^{\mu \nu} (p_1, \cdots, p_n;  q)= - \frac{1}{\sqrt{Z}} \sum_{k=0}^{\infty}  \frac{(2k)!}{2^{2k} (k!)^2  }F^{ -4(k+1) }  \int d^4 x\, e^{-i q \cdot x} \qquad\qquad\non\\
 \times \<0|   \left\{ \pi \left[ \partial^\mu \pi \partial^\nu \pi - \frac{2}{k+1}  g^{\mu \nu} (\partial \pi)^2 \right] (\partial \pi)^{2k} \right\} (x) | \pi (p_1) \cdots \pi (p_n)\>.\quad\label{eq:dbires}
\eea
Again, as the integrand on the RHS of Eq. (\ref{eq:dbires}) has at least three fields in it, we have the regularity condition $ \lim_{q\to 0} q_\mu  r^{\mu \nu} (p_1, \cdots, p_n;  q) = 0$. Therefore, for  momentum $q$ to be soft, we see that the on-shell amplitudes of DBI scalars display enhanced soft behavior:
\bea
M(p_1, \cdots, p_n, \tau q) = \tau^2 q_\mu q_\nu r^{\mu \nu} (p_1, \cdots, p_n;  0)+ \ordr(\tau^3).
\eea

\begin{figure}[htbp]
\begin{center}
\includegraphics[height=6cm]{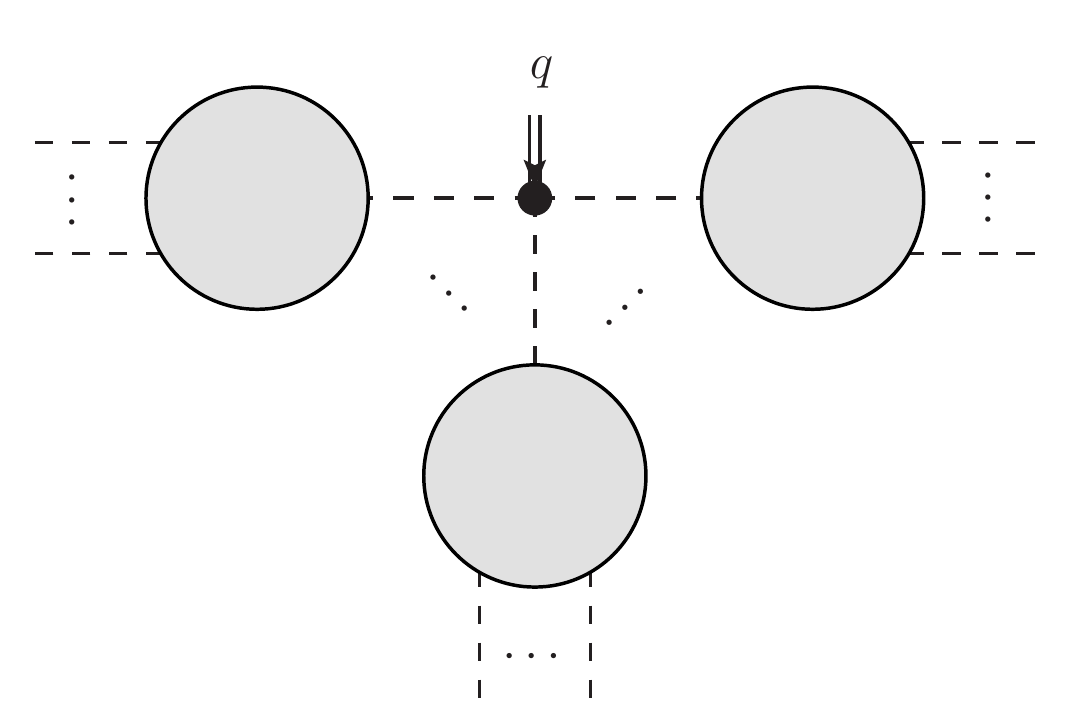}
\end{center}
\caption{\label{fig:rcj} The terms in the remainder function $r^{\mu \nu} (p_1, \cdots, p_n;  q)$ of DBI at tree level. The $(2n+1)$ point vertex with momentum insertion $q$ is connected with semi-on-shell sub-amplitudes, which are represented by blobs.}
\end{figure}
As we mentioned earlier in Sec. \ref{sec:nlsm}, at tree level the relation for the semi-on-shell amplitude becomes a Berends-Giele recursion relation. The integral of $x$ on the RHS of Eq. (\ref{eq:dbires}) gives the following $(2n+1)$-point vertex
\bea
\nV^{\dbi}_{2k+1} (p_1, \cdots, p_{2k+1}) &=& i \frac{(2k-2)!}{2^{2k-2} [(k-1)!]^2  } \left(-F^{-4}\right)^k \non\\
&&\times \sum_{\sigma \in S_{2k+1}} \left[  (q\cdot p_{\sigma (1)}) (q \cdot p_{\sigma (2)}) \prod_{i=1}^{k-1} p_{\sigma( 2i+1)} \cdot p_{\sigma (2i+2) }\right.\non\\
&&\left.  -\frac{2}{k}  q^2 \prod_{i = 1}^k p_{\sigma (2i-1)} \cdot p_{\sigma (2i)}  \right],
\eea
where $q = -\sum_{i=1}^{2k+1} p_i$, and $\sigma$ is a permutation of $\{1, 2, \cdots, 2k+1 \}$.  At tree level, Eq. (\ref{eq:dbires}) can be seen as these vertices connected with semi-on-shell sub-amplitudes, as shown in Fig. \ref{fig:rcj}. Then Eq. (\ref{eq:dbisaq}) becomes the Berends-Giele recursion relation
\bea
J (p_1, \cdots, p_n) =\frac{i}{q^2} \sum_{k=1}^{[n/2]}  \sum_{l} \nV^{\dbi}_{2k+1} (q_{l^1}, \cdots, q_{l^{2k
+1}}) \prod_{i=1}^{2k+1} J( \{ p_{l^i_j} \}),\label{eq:dbibg}
\eea
where $l$ is a  splitting of the non-ordered set $\{1,2,\cdots, n\}$ into $2k+1$ disjoint non-ordered subsets $l^1, l^2, \cdots , l^{2k+1}$, and $q_{l^{m}} = \sum_{i \in l^m} p_i$. Notice that once we have the Feynman rules, we can directly derive a version of the Berends-Giele recursion relation, which is different from the above. The relation is not unique because it relates off-shell objects, which rely on Feynman rules that can change under field redefinition. In our relation, the vertices $\nV^\dbi_{2k+1}$ explicitly contains two powers of the off-shell leg, which is an indication of the enhanced soft behavior. Taking the on-shell limit and soft limit for $q$, Eq. (\ref{eq:dbibg}) becomes
\bea
M (p_1, \cdots, p_n, \tau q) &=& \tau^2 q_\mu q_\nu \sum_{k=1}^{[n/2]}  \sum_{l} \frac{(2k-2)!}{2^{2k-2} [(k-1)!]^2  }  \left(-F^{-4}\right)^k\non\\
&&\times\sum_{\sigma \in S_{2k+1}} \left[   q_{l^{\sigma(1)}}^\mu \ q_{l^{\sigma(2)}}^\nu \prod_{m=1}^{k-1} q_{l^{\sigma(2m+1)}} \cdot q_{l^{\sigma(2m+2)}} \right]  \prod_{i=1}^{2k+1} J( \{ p_{l^i_j} \}),\label{eq:dbisstl}
\eea
which is the tree level result for the subleading single soft theorem of DBI. Notice that in Eq. (\ref{eq:dbisstl}) the terms inside $\nV^\dbi_{2k+1}$ that are proportional to $q^2$ have been dropped.

\subsection{Galileon theory}
\label{secggal}

The Galilean symmetry for scalar theories is defined as the invariance under the shift
\bea
\pi (x)  \to \pi (x) + \vep + a_\mu x^\mu,\label{eq:ggals}
\eea
where $\vep $ and $a_\mu$ are constants. In the Lagrangian, terms with enough powers of derivatives, for example functions of $ \partial^\mu  \partial^\nu \pi$, trivially realize the symmetry. However, there are certain non-trivial terms in the Lagrangian with less than two derivatives per field that preserve the shift symmetry. It turns out that for a certain power of $\pi$ field, this kind of non-trivial term can be fixed up to the overall normalization and total derivatives. The theory with a Lagrangian that completely consists of these non-trivial terms is called the Galileon theory \cite{Nicolis:2008in}, and the scalars in the theory are known as Galileons. Originally proposed as a modification of gravity, the Galileon theory has been studied in various context, including the soft behavior of its scattering amplitudes.

There are many  ways to write the terms in the Galileon Lagrangian, which are equivalent up to total derivatives. The useful properties of these terms are reviewed in Appendix \ref{app:gall}. The form that is particularly convenient to use for our purpose is
\bea
\lag_\gal = - \frac{1}{2} \pi \sum_{n=1}^d c_{n+1} \lag_n^\TD,\label{eq:ggallag}
\eea
where $c_n$ are constant coefficients, and we need $c_2 =1$  to have a canonically normalized kinetic term. The term $\lag_n^\TD$ is defined as
\begin{align}
\lag_n^{\TD} \equiv \Pi_{\mu \nu} T^{\mu \nu}_n, \qquad   T^{\mu \nu}_n \equiv \sum_{\sigma \in S_n} \epsilon (\sigma) g^{\mu}_{ \mu_1} g^{\nu \mu_{\sigma (1)}} \prod_{i=2}^n \Pi^{\mu_{\sigma (i)}}_{\mu_i},\label{eq:galltdd}
\end{align}
where $\Pi^{\mu \nu} \equiv \partial^\mu  \partial^\nu \pi$, and $\epsilon (\sigma)$ is the signature of the permutation $\sigma$.

We will work with $d=4$. The first thing to inspect is the Ward identity under the constant shift $\pi \to \pi + \vep$. When we promote the constant $\vep$ to $\vep(x)$, the change in the Lagrangian is
\bea
\delta \lag_\gal = \vep (x) \partial_\mu \J_\gal^\mu (x)
\eea
up to total derivatives, with
\bea
\partial_\mu \J^\mu_\gal  =\DAl \pi + \frac{1}{2} \partial_\mu \partial_\nu \left[\pi \left(  3c_3 T^{\mu \nu}_2 + 4c_4 T^{\mu \nu}_3 + 5c_5 T^{\mu \nu}_4\right) \right].\label{eqwiggal}
\eea
Then the corresponding Ward identity is
\bea
i\<\partial_\mu \J^\mu_\gal(x)  \prod_{i=1}^n \pi (x_i) \> =  \sum_{r=1}^n \< \pi (x_1) \cdots  \delta^{(4)}(x-x_r) \cdots \pi (x_n) \> .\label{eqggalwi}
\eea

\begin{figure}[htbp]
\centering
\includegraphics[height=3cm]{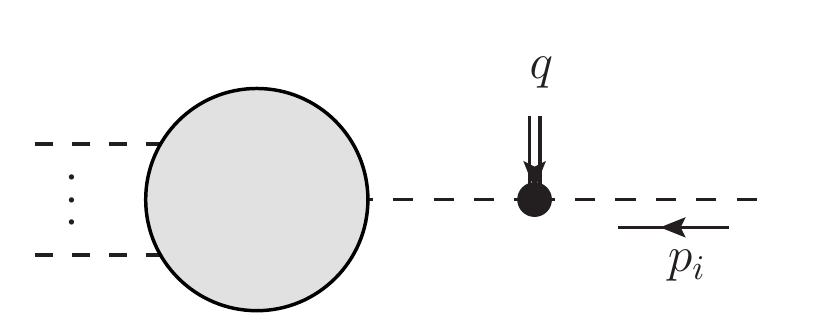}
\caption{\label{fig:poleinr} The ``pole diagram'' in the remainder function $R^{\mu \nu}_\gal$ when cubic terms exist in the Lagrangian.}
\end{figure}
We have two powers of total derivatives in Eq. (\ref{eqwiggal}), but this does not necessarily mean that the amplitudes will have the enhanced Adler's zero. Actually, the term with coefficient $c_3$ in Eq. (\ref{eqwiggal}) has two powers of $\pi$, which generates a 2-pt vertex with soft momentum injection. When we calculate the remainder function $R^{\mu \nu}_\gal$, we will have the so-called ``pole diagrams'' as shown in Fig. \ref{fig:poleinr}, where the 2-pt vertex is connected to an external leg with momentum $p_i$. Then the propagator connected to this vertex, $i/(2\,p_i \cdot q)$, will generate a pole in the soft momentum $q$. Therefore, the regularity condition $\lim_{q \to 0} q_\mu R^{\mu \nu}_\gal = 0$ is no longer satisfied. We can see that it is exactly the cubic term in the Lagrangian with the coefficient $c_3$ that spoils the regularity condition. However, due to Galileon duality \cite{Fasiello:2013woa,deRham:2013hsa,Kampf:2014rka}, $c_3$ can always be set to $0$ with a field redefinition. This indicates that when $c_3 \ne 0$, there are multiple pole diagrams and their contribution in the leading order of the soft momentum must cancel each other in the on-shell amplitude. Such a condition can be viewed as a constraint on the specific form of the Galileon interactions. For simplicity we can just set $c_3 = 0$. By performing LSZ reduction to the Ward identity given by Eq. (\ref{eqggalwi}), we can express the semi-on-shell amplitude of the Galileon theory as
\bea
q^2 J (p_1, \cdots ,p_n)   =-\sqrt{Z}  q_\mu q_\nu R^{\mu \nu}_\gal (p_1, \cdots, p_n; q),\label{eq:ggaljg}
\eea
where $q = - \sum_{i=1}^n p_i$. The remainder function is given by
\bea
R^{\mu \nu}_\gal (p_1, \cdots, p_n; q) = \frac{1}{\sqrt{Z}} \int d^4 x\, e^{-i q \cdot x}  \<0|   \frac{1}{2} \left[ \pi \left(   4c_4 T^{\mu \nu}_3 + 5c_5 T^{\mu \nu}_4\right) \right] (x) | \pi (p_1) \cdots \pi (p_n) \>,\quad\quad\label{eqggalqsos}
\eea
which satisfies the regularity condition. Taking the on-shell and soft limit for momentum $q$, we arrive at the soft theorem of the on-shell amplitude in the Galileon theory:
\bea
M(p_1, \cdots, p_n, \tau q) &=&  \tau^2 q_\mu q_\nu R^{\mu \nu}_\gal (p_1, \cdots, p_n; 0)+ \ordr(\tau^3),\label{eqggalqos}
\eea
which shows the enhanced soft behavior explicitly and also gives the subleading single soft limit.

Similar to Section \ref{sec:dbi}, we can generate Berends-Giele recursion relations at tree level. The terms in Eq. (\ref{eqggalqsos}) with coefficients $c_4$ and $c_5$ give the 3- and 4-pt vertices:
\bea
\nV^\gal_3 (p_1, p_2, p_3) &=& 2ic_4 \sum_{\sigma \in S_3} \left\{q^2 \left[ p_{\sigma (1)}^2 p_{\sigma(2)}^2 - p_{\sigma(1), \sigma(2)}^2  \right] -2 p_{q, \sigma(1)}^2 p_{\sigma(2)}^2 \right. \non\\
&&\left.+ 2 p_{q,\sigma(1)} p_{q,\sigma(2)} p_{\sigma(1),\sigma(2)}  \right\},\\
\nV^\gal_4 (p_1, p_2, p_3,p_4) &=&-\frac{5}{2} ic_5\sum_{\sigma \in S_4}  \left\{ q^2 \left[ p_{\sigma(1)}^2 p_{\sigma(2)}^2 p_{\sigma(3)}^2 - 3 p_{\sigma(1)}^2 p_{\sigma(1),\sigma(2)}^2 \right.\right.\non\\
&&\left.+ 2 p_{\sigma(1),\sigma(2)} p_{\sigma(2),\sigma(3)} p_{\sigma(1),\sigma(3)} \right] -3 p_{q,\sigma(1)}^2 \left[p_{\sigma(2)}^2 p_{\sigma(3)}^2 - p_{\sigma(2),\sigma(3)}^2 \right] \nonumber\\
&&\left. + 6 p_{q,\sigma(1)} p_{q,\sigma(2)}\left[ p_{\sigma(1),\sigma(2)} p_{\sigma(3)}^2 - p_{\sigma(1),\sigma(3)} p_{\sigma(2),\sigma(3)} \right]  \right\},
\eea
where $q \equiv - \sum_{i=1}^n p_i$ in vertex $\nV_n^\gal$, $p_{i,j} \equiv p_i \cdot p_j$ and $p_{q,i} \equiv q \cdot p_i$. At tree level, Eq. (\ref{eq:ggaljg}) leads to the Berends-Giele recursion relation:
\bea
J (p_1, \cdots, p_n) &=&  \frac{i}{ q^2}  \sum_{l} \nV^\gal_3 (q_{l^1}, q_{l^2}, q_{l^3}) \prod_{i=1}^3 J ( \{ p_{l^i_j} \})\nonumber\\
&&+\frac{i}{ q^2}  \sum_{l} \nV^\gal_4 (q_{l^1}, q_{l^2}, q_{l^3},q_{l^4}) \prod_{i=1}^4 J ( \{ p_{l^i_j} \})\label{eqggaltlbg},
\eea
and the single soft theorem at tree level is
\bea
M(p_1, \cdots, p_n, \tau q) =\tau^2 \sum_{l} 4c_4 \sum_{\sigma \in S_3} \left\{  p_{q, l^{\sigma(1)}} p_{q, l^{\sigma(2)}} p_{l^{\sigma(1)}, l^{\sigma(2)}} - p_{q, l^{\sigma(1)}}^2 q_{l^{\sigma(2)}}^2 \right\} \prod_{i=1}^3 J ( \{ p_{l^i_j} \})\nonumber\\
 +\tau^2 \sum_{l} \frac{15}{2} c_5 \sum_{\sigma \in S_4} \left\{  p_{q, l^{\sigma(1)}}^2 \left[q_{l^{\sigma(2)}}^2 q_{l^{\sigma(3)}}^2 - p_{l^{\sigma(2)}, l^{\sigma(3)}}^2 \right] \right. \nonumber\\
 \left. - 2 p_{q, l^{\sigma(1)}} p_{q, l^{\sigma(2)}}\left[ p_{l^{\sigma(1)}, l^{\sigma(2)}} q_{l^{\sigma(3)}}^2 - p^{\phantom{2}}_{l^{\sigma(1)}, l^{\sigma(3)}} p_{l^{\sigma(2)}, l^{\sigma(3)}} \right]  \right\}  \prod_{i=1}^4 J ( \{ p_{l^i_j} \}),\ \label{eqggaltlsf}
\eea
with $p_{l^i,j} \equiv q_{l^i} \cdot p_j$ and $p_{l^i,l^j} \equiv q_{l^i} \cdot q_{l^j}$.

The Ward identity corresponding to the full shift symmetry will not give any additional information. As discussed in Section \ref{sec:dbi}, we can derive a relation between the current $\J_\gal^\mu$ of the constant shift and the current $j_\gal^{(\mu) \nu}$ of the enhanced shift $\pi \to \pi+ a_\mu x^\mu$:
\begin{align}
\partial_{\nu}  j_\gal^{ ( \mu ) \nu} &= x^\mu \partial_{\nu} \J_\gal^\mu,\label{eqggalesj}
\end{align}
and the RHS of Eq. (\ref{eqggalesj}) can indeed be written as a total derivative. Then the Ward identity under the enhanced shift symmetry is
\bea
i\<x^\nu \partial_\mu \J^\mu_\gal(x)  \prod_{i=1}^n \pi (x_i) \> =  \sum_{r=1}^n \< \pi (x_1) \cdots  \delta^{(4)}(x-x_r) x^\nu \cdots \pi (x_n) \>,
\eea
which is just Eq. (\ref{eqggalwi}) multiplied by $x^\nu$. Therefore, Eq. (\ref{eqggalwi}) contains all the information given by the shift symmetry, and as we have shown, it is enough to guarantee the enhanced soft behavior.  In general, the Ward identity corresponding to the constant shift is enough for us to see the soft degree $s$ in the theory, unless the enhanced shift has a field dependent component, like that of DBI given by Eq. (\ref{eq:dbis}). Such a field dependent shift is exactly a feature of exceptional field theories. As classified in Ref. \cite{Cheung:2016drk}, although the soft degree of the ordinary Galileon theory is non-trivial, it is not exceptional.

\subsection{Special Galileon}

The special Galileon is a special case of the Galileon theory, whose amplitudes vanish faster than the ordinary Galileon theory. Discovered nearly at the same time in different contexts \cite{Cachazo:2014xea,Cheung:2014dqa}, the hidden symmetry protecting the exceptional soft behavior is soon found to be the symmetry under the enhanced shift \cite{Hinterbichler:2015pqa}
\bea
\pi \to \pi + \theta^{\mu \nu} (\alpha^2 x_{\mu} x_{\nu} + \partial_\mu \pi \partial_\nu \pi),\label{eq:sgals}
\eea
where $\theta_{\mu \nu}$ is a constant traceless tensor, namely $\theta_{\mu \nu} g^{\mu \nu} = 0$. In 4 dimensions, the Lagrangian of special Galileon is given by setting $c_3 = c_5 = 0$ and $c_4 =- 1/(12\alpha^2)$ in the Galileon Lagrangian. From Eq. (\ref{eq:ggallag}) we know that
\bea
\lag_\sgal =- \frac{1}{2} \pi \left(\lag^\TD_1 -\frac{1}{12 \alpha^2}  \lag^\TD_3  \right).
\eea
As a special case of the ordinary Galileon theory, we can readily use the results in Section \ref{secggal}. For example, the Ward identity for special Galileon under the constant shift $\pi \to \pi + \vep$ is
\bea
i  \partial_\mu \< \J_\sgal^\mu (x) \prod_{i=1}^n \pi (x_i) \>  = \sum_{r=1}^n \<\pi (x_1) \cdots    \delta^{(4)}(x -x_r)  \cdots \pi (x_n) \>,\label{eqsgalwics}
\eea
where
\bea
\partial_\mu \J_\sgal^\mu =\DAl \pi -  \frac{1}{6 \alpha^2} \partial_\mu \partial_\nu \left(\pi T^{\mu \nu}_3\right) .
\eea
Eq. (\ref{eqsgalwics}) leads to the same soft behavior given by Section \ref{secggal}, namely Eq. (\ref{eqggalqos}). On surface, these results indicate that the leading single soft behavior of special Galileon amplitudes is at least at $\ordr( \tau^2)$, where $q$ is the soft momentum. However,  the leading single soft contribution of special Galileon is actually at $\ordr(\tau^3)$. As we discussed in Sec. \ref{secggal}, the shift given by Eq. (\ref{eq:sgals}) is field dependent, thus the corresponding Ward identity contains more information than  Eq. (\ref{eqsgalwics}), which is a feature of exceptional theories classified in Ref.  \cite{Cheung:2016drk}.

Again, the current $j^{\{\mu \nu\} }_\rho$ corresponding to the enhanced shift given by Eq. (\ref{eq:sgals}) is related to the current $\J^\mu_\sgal$ of the constant shift symmetry:
\bea
\partial^{\rho}  j^{\{\mu \nu\}}_{ \rho} &=\left( \alpha^2 x^{\{ \mu} x^{ \nu \} } + \partial^{\{\mu } \pi \partial^{\nu \} } \pi \right) \partial_\rho \J^\rho_\sgal ,\label{eqchmf}
\eea
where the braces $\{\cdots\}$ for the Lorentz indices denote symmetric and traceless operations, namely $a_{\{ \mu} b_{\nu \}} \equiv (a_\mu b_\nu + a_\nu b_\mu)/2 - g_{\mu \nu} a\cdot b/4$. We prove in Appendix \ref{app:gall} that the RHS of Eq. (\ref{eqchmf}) can indeed be written as a total derivative. The Ward identity for special Galileon under the enhanced shift symmetry is:
\bea
&&i\< \left( \alpha^2 x^{\{ \mu} x^{ \nu \} } + \partial^{\{\mu } \pi \partial^{\nu \} } \pi \right) \partial_\rho \mathcal{J}^\rho_\sgal  \prod_{i=1}^n \pi (x_i) \> \non\\
&=& \sum_{r=1}^n \< \pi  (x_1) \cdots  [ \alpha^2 x^{\{ \mu} x^{ \nu \} } + \partial^{\{\mu } \pi (x) \partial^{\nu \} } \pi (x) ] \delta^{(4)}(x-x_r) \cdots \pi (x_n) \>.\label{eqsgalwiin}
\eea
Again, the terms proportional to $x^{\{ \mu} x^{\nu \}}$ on both sides of Eq. (\ref{eqsgalwiin}) cancel out  as a consequence of Eq. (\ref{eqsgalwics}). Adding two powers of total derivatives to contract the free Lorentz indices in Eq. (\ref{eqsgalwiin}), we arrive at
\bea
&&i \<\left[ \DAl (\pi + A) + \partial_\mu \partial_\nu \partial_\rho B^{\mu \nu \rho} +  \partial_\mu \partial_\nu \partial_\rho \partial_\lambda C^{\mu \nu \rho \lambda}\right](x) \prod_{i=1}^n \pi (x_i) \>\non\\
&=& \sum_{r=1}^n \<\pi (x_1) \cdots  \left\{  \delta^{(4)}(x -x_r)  + \partial_\mu \partial_\nu \left[ \frac{1}{2 \alpha^2} \partial^{\{ \mu} \pi (x) \partial^{\nu \}} \pi (x) \delta^{(4)}(x -x_r) \right] \right\} \cdots \pi (x_n) \>,\quad\quad\label{eqsgalwifi}
\eea
where
\bea
A &=& \frac{1}{12 \alpha^2} \pi \lag_2^\TD - \frac{1}{4 \alpha^2} \partial_\rho \left( \pi \partial^\rho \pi \lag_1^\TD \right) + \frac{1}{40 \alpha^4} \partial_\rho \left( \pi \partial^\rho \pi \lag_3^\TD \right),\\
B^{\mu \nu \rho} &=& -\frac{1}{ 3 \alpha^2}  \partial^{\{ \mu} \pi \partial^{\nu \}} \pi \partial_\lambda \pi T_1^{\rho \lambda} + \frac{1}{10 \alpha^4}  \partial^{\{ \mu} \pi \partial^{\nu \}} \pi \partial_\lambda \pi T_3^{\rho \lambda},\\
C^{\mu \nu \rho \lambda} &=&  \frac{1}{6 \alpha^2} \pi \partial^{\{ \mu} \pi \partial^{\nu \}} \pi T_1^{\rho \lambda} - \frac{1}{20 \alpha^4} \pi \partial^{\{ \mu} \pi \partial^{\nu \}} \pi T_3^{\rho \lambda}.
\eea
Detailed derivation is presented in Appendix \ref{app:gall}. Because of the existence of the first term in $A$, the LHS of Eq. (\ref{eqsgalwifi}) can never be written as $\DAl \pi$ added by a term with 3 powers of total derivatives. However, the term $A$ will vanish when we take the soft momentum on-shell because of the d'Alembert operator in front of it. On the other hand, the $C$ term will give 4 powers of soft momentum, thus the only relevant term for the leading soft behavior at $\ordr(\tau^3)$ is $B$.

Eq. (\ref{eqsgalwifi}) leads to the following expression for the semi-on-shell amplitude of the special Galileon:
\bea
q^2 J(p_1, \cdots, p_n) &=& -q^2  \int d^4 x\, e^{-i q \cdot x}  \<0|   A(x) | \pi (p_1) \cdots \pi (p_n)\>\non\\
&&-iq_{\mu } q_{\nu} q_{\rho} \int d^4 x\, e^{-i q \cdot x}  \<0|   B^{\mu \nu \rho} (x) | \pi (p_1) \cdots \pi (p_n)\>\non\\
&&+q_{\mu } q_{\nu} q_{\rho} q_{\lambda} \int d^4 x\, e^{-i q \cdot x}  \<0|  C^{\mu \nu \rho \lambda} (x) | \pi (p_1) \cdots \pi (p_n)\>.\label{eqsgalsosqr}
\eea
Taking $q$ to be on-shell and soft, we arrive at the enhanced soft behavior of the on-shell amplitude of the special Galileon:
\bea
M(p_1, \cdots, p_n, \tau q) &=& \tau^3 \frac{i}{\sqrt{Z}}   q_{\mu} q_{\nu} q_{\rho} \int d^4 x\,    \<0|  B^{\mu \nu \rho} (x) | \pi (p_1) \cdots \pi (p_n)\> + \ordr(\tau^4).
\eea
At tree level, Eq. (\ref{eqsgalsosqr}) leads to the Berends-Giele recursion relation
\bea
J (p_1, \cdots, p_n) &=&  \frac{i}{ q^2}  \sum_{l} \nV^\sgal_3 (q_{l^1}, q_{l^2}, q_{l^3}) \prod_{i=1}^3 J ( \{ p_{l^i_j} \})\nonumber\\
&&+\frac{i}{ q^2}  \sum_{l} \nV^\sgal_5 (q_{l^1}, \cdots,q_{l^5}) \prod_{i=1}^5 J ( \{ p_{l^i_j} \}),
\eea
where
\bea
\nV^\sgal_3 (p_1, p_2, p_3) &=& -i\frac{1}{24 \alpha^2} \sum_{\sigma \in S_3} \left\{2q^2 \left[ p_{\sigma (1)}^2 p_{\sigma(2)}^2 - p_{\sigma(1), \sigma(2)}^2 +p_{\sigma(1), \sigma(2)} p_{q,\sigma(3)} \right]\right. \non\\
&& -6 q^2 p_{q, \sigma(1)} p_{\sigma(2)}^2  - 8 p_{q,\sigma(1)} p_{q,\sigma(2)} p_{q,\sigma(3)} - 4 q^2 p_{q,\sigma(1)} p_{q,\sigma(2)}\non\\
&&\left. + q^4 p_{\sigma(1), \sigma(2)} \right\},\\
\nV^\sgal_5 (p_1, p_2, p_3,p_4, p_5) &=&-i \frac{1}{80 \alpha^4} \sum_{\sigma \in S_5}  \left\{2 q^2 p_{q, \sigma (4)} \left[ p_{\sigma(1)}^2 p_{\sigma(2)}^2 p_{\sigma(3)}^2 - 3 p_{\sigma(1)}^2 p_{\sigma(1),\sigma(2)}^2 \right.\right.\non\\
&&\left.+ 2 p_{\sigma(1),\sigma(2)} p_{\sigma(2),\sigma(3)} p_{\sigma(1),\sigma(3)} \right] \non\\
&&+ \left[8 p_{q,\sigma(1)} p_{q,\sigma(2)} - 2 q^2 p_{\sigma(1), \sigma(2)} \right] \left( p_{q,\sigma(3)} \left[p_{\sigma(4)}^2 p_{\sigma(5)}^2 - p_{\sigma(4),\sigma(5)}^2 \right]  \right.\nonumber\\
&&\left. - 2 p_{\sigma (3), \sigma(4)} \left[ p_{q,\sigma(4)} p_{\sigma(5)}^2 - p_{\sigma(4),\sigma(5)} p_{q,\sigma(3)} \right] \right)\non\\
&&+  \left[4 p_{q,\sigma(1)} p_{q,\sigma(2)} - q^2 p_{\sigma(1), \sigma(2)} \right] \left( q^2 \left[p_{\sigma(3)}^2 p_{\sigma(4)}^2 - p_{\sigma(3),\sigma(4)}^2 \right]  \right.\non\\
&&\left.\left.-2 \left[ p^2_{q,\sigma(3)} p_{\sigma(4)}^2 - p^2_{q,\sigma(3)} p_{q,\sigma(4)} p_{\sigma(3),\sigma(4)}\right] \right) \right\}.
\eea
The leading soft behavior for the on-shell amplitude can be much simplified as we only keep $\ordr (\tau^3) $ terms and drop the terms with a factor of $q^2$:
\bea
M(p_1, \cdots, p_n, \tau q) &=& \tau^3 \left\{ \sum_{l} \frac{2}{\alpha^2}  p_{q, l^{1}} p_{q, l^{2}} p_{q, l^{3}} \prod_{i=1}^3 J ( \{ p_{l^i_j} \})\right.\nonumber\\
&&\left.- \sum_{l} \frac{1}{6 \alpha^4}  \sum_{\sigma \in S_5}   p_{q, l^{\sigma(1)}} p_{q, l^{\sigma(2)}} p_{q, l^{\sigma(3)}} \left[q_{l^{\sigma(4)}}^2 q_{l^{\sigma(5)}}^2 - p_{l^{\sigma(4)}, l^{\sigma(5)}}^2 \right] \prod_{i=1}^5 J ( \{ p_{l^i_j} \})\right\}\nonumber\\
&& + \ordr (\tau^4).\label{eqsgalsslf}
\eea
Like before, here $q$ is the soft momentum, while $ q_{l^{m}} \equiv \sum_{i \in l^m} p_i$ is the  sum of a subset of hard external momenta, $p_{l^i,q} \equiv q_{l^i} \cdot q$ and $p_{l^i,l^j} \equiv q_{l^i} \cdot q_{l^j}$.

\section{The extended theories}
\label{sec:ext}

In the previous section, we have expressed the subleading single soft limit of tree level amplitudes using semi-on-shell sub-amplitudes, as seen in Eqs. (\ref{eq:dbisstl}), (\ref{eqggaltlsf}) and (\ref{eqsgalsslf}). The subleading single soft theorems of scalar EFTs are first studied in Ref. \cite{Cachazo:2016njl} using the CHY representation of scattering amplitudes. Interestingly, for NLSM and special Galileon, the subleading single soft factor of the theory was expressed as on-shell amplitudes of a different theory, which involves more than one kind of scalars interacting with each other. These theories with mixed scalars were seen as extended version of the original theory emerging from the soft limit. The CHY representation of these extended theories is straightforward, but only gives information of the on-shell amplitudes, while the details of the interaction vertices and the Lagrangian that generate these vertices are hidden. It is not until Ref. \cite{Low:2017mlh,Low:2018acv} presented the subleading single soft limit of NLSM, as discussed in Section \ref{sec:nlsm}, that the Feynman vertices of the corresponding extended theory are identified. The method of using Ward identities to generate soft theorems is complementary to the approach of Ref. \cite{Cachazo:2016njl}, which helps us take a step further to understanding the nature of the extended theories.

In the following, we will first review the identification of  Feynman rules for the extended theory of NLSM. Drawing inspiration from the previous success, we then proceed to identify the interaction vertices of the extended theory of special Galileon, which exhibits a richer structure. Finally, we write down the Lagrangian for the extension of special Galileon.

\subsection{Extension of the Nonlinear Sigma Model}
\label{sec:nlsme}

The NLSM amplitudes considered in Ref. \cite{Cachazo:2016njl} is flavor-ordered; namely, the flavor structure of the on-shell amplitudes is stripped:\footnote{The normalization of generators in this work is given by $\tr \{ X^a X^b\} = \delta^{ab}$, $\tr \{X^a T^i\} = 0$, and $\tr \{T^i T^j \} = \delta^{ij}$.}
\bea
M^{a_1 \cdots a_{n} } (p_1, \cdots, p_{n}) \equiv \sum_{\sigma \in S_{n-1}} {\rm Tr} ( X^{a_n} X^{a_{\sigma (1)}} \cdots X^{a_{\sigma (n-1)}} ) M_\sigma(p_1,\cdots,p_{n-1})\ ,
\eea
where $M_\sigma(p_1,\cdots,p_{n-1}) \equiv M(\sigma)$ is the flavor ordered amplitude, and $X^a$ is the aforementioned broken generator. Flavor-ordering for amplitudes, as reviewed in Appendix \ref{app:flavo}, is more useful when two flavor traces can be merged into a single trace. This can be done in $SU(N)$ NLSM \cite{Kampf:2013vha}, which we will consider below. At tree level and after flavor ordering, the Ward identity for on-shell amplitudes of NLSM given by Eq. (\ref{eqnlsmssql}) becomes
\bea
M^{\text{NLSM}}_{n+1} (\mathbb{I}_{n+1}) &=& \sum_{k=1}^{[n/2]}  \frac{-(-4)^k}{(2k+1)!f^{2k}} \sum_{l}\sum_{j=1}^{2k-1} \left[ \left(\begin{array}{c}
2k\\
j
\end{array} \right) (-1)^{j} -1 \right] p_{n+1} \cdot q_{l_{j+1}}\non\\
&&\times \prod_{m=1}^{2k+1} J (l_{m-1}+1, \cdots , l_m)\ ,\label{eqnlsmosex}
\eea
where $\mathbb{I}_{n+1} = \{1,2, \cdots, n+1\}$ is the identity permutation. In the above $l$ is a way to split $\{1, 2, \cdots,n\}$ into $2k+1$ disjoint, ordered subsets $\{l_{m-1}+1, \cdots , l_m \}$, with $l_0=0$, $l_{2k+1} = n$ and $q_{l_{j+1}} = \sum_{i=l_j+1}^{l_{j+1}} p_i$. We have expressed the on-shell amplitude using flavor-ordered semi-on-shell sub-amplitudes $J(\sigma)$, and as it is proportional to $p_{n+1}$, the Adler's zero is manifest when we take $p_{n+1} \to 0$.

The subleading single soft theorem derived using the CHY representation, on the other hand, is
\bea
M^{\text{NLSM}}_{n+1}(\mathbb{I}_{n+1}) = \frac{\tau}{\lambda f^2} \sum_{i=2}^{n-1}  s_{n+1,i} \ M_n^{\text{NLSM}\oplus \phi^3}(\mathbb{I}_{n}|1,n,i) + \ordr (\tau^2), \label{eqchynlsm}
\eea
where we have taken the soft limit for $p_{n+1}$, and $s_{i,j} \equiv (p_i+ p_j)^2$. A review of the CHY representations of various scalar EFTs is presented in Appendix \ref{app:chyr}. Compared with Ref. \cite{Cachazo:2016njl}, we have added the additional $1/(\lambda f^2)$ factor to match the mass dimensions on both sides of the equation, as a consequence of keeping the coupling constants manifest in the amplitudes. The RHS of the above equation contains the amplitude of the extended theory denoted by $\text{NLSM}\oplus \phi^3$, which involves both the NGB $\Sigma^a$ and the biadjoint scalar $\phi^{a\tilde{a}}$.\footnote{Before this section we follow the common convention of always using $\pi$  to represent the single scalar in a ``pure'' theory like NLSM, DBI, and Galileon. In the extended theories we need to discern the difference between NGBs and Galileons, so in the following we will use the convention in Ref. \cite{Cachazo:2016njl} to denote NGBs as $\gs$, and reserve $\pi$ exclusively for Galileons.} Such a theory can also arise in the low-energy limit of Z-theory \cite{Carrasco:2016ygv}. The field $\phi^{a\tilde{a}}$ furnishes the adjoint representations of two copies of $SU(N)$ group, so that the index $a$ refers to the $SU(N)$ under which the NGBs are charged, while $\tilde{a}$ refers to the second copy of $SU(N)$, which we will denote as $SU(\tilde{N})$. The self-interaction of $\phi^{a \tilde{a}}$ is given by a single cubic term in the Lagrangian:
\bea
-\frac{\lambda}{6} \phi^{a \ta} \phi^{b \tb} \phi^{c \tc} f^{abc} f^{\ta \tb \tc}.\label{eq:phiself}
\eea
Biadjoint scalars have been discovered to have a simple CHY representation \cite{Cachazo:2013iea}, and can act as a building block for gauge theory and gravity amplitudes \cite{BjerrumBohr:2012mg}. In the extended theory, $(\alpha| \beta)$ is used to represent the flavor ordering of $\alpha$ for $SU(N)$ and $\beta$ for $SU(\tilde{N})$, so that $(\mathbb{I}_{n}|1,n,i)$ in Eq. (\ref{eqchynlsm}) denotes amplitudes with $n$ scalars, with the external fields of labels $1$, $n$ and $i$ to be $\phi$ while the other fields are normal NGBs of $SU(N)$.

Comparing Eq. (\ref{eqnlsmosex}) with Eq. (\ref{eqchynlsm}), we arrive at the following three observations about Feynman rules in $\nlsm \oplus \phi^3$ \cite{Low:2017mlh,Low:2018acv}:
\begin{enumerate}
\item $SU(\tilde{N})$ charge conservation: No vertex exists with only one $\phi$.
\item Soft constraint on even-pt amplitudes: Vertices with two $\phi$ are the same as vertices of the pure NLSM with the same ordering of $SU(N)$ indices. This implies the absence of odd-pt vertices with two $\phi$.
\item Structure of the current entering the odd-pt amplitudes: Odd-pt vertices with three $\phi$, two of which have $SU(N)$ indices adjacent to each other, are given by
\bea
V_{2k+1}^{\text{NLSM} \oplus \phi^3} (\mathbb{I}_{2k+1} | 1,2k+1,j)  =\frac{i}{2} \frac{-(-4)^k \lambda }{(2k+1)!f^{2k-2}} \left[ \left(\begin{array}{c}
2k\\
j-1
\end{array} \right) (-1)^{j-1} -1 \right]\ . \label{eqmixedfr}
\eea
\end{enumerate}
The first observation comes from the fact that $\phi^{a\tilde{a}}$ furnishes the adjoint representation of $SU(\tilde{N})$, so the interactions should conserve the $SU(\tilde{N})$ charge. At the level of the Lagrangian, it is impossible to construct an interaction term with only one $\phi$ carrying an $SU(\tilde{N})$ index: such a term is not a flavor scalar. This observation then implies that the $SU(\tilde{N})$ flavor flow must be continuous in a Feynman diagram. For the tree amplitude $M_n^{\text{NLSM}\oplus \phi^3}(\mathbb{I}_{n}|1,n,i)$, we have 3 external $\phi$ legs. Therefore, in a certain diagram, there cannot be any vertices with more than three $SU(\tilde{N})$ indices. Moreover, there must be exactly one vertex with three $\phi$, while a number of vertices with two $\phi$ are needed to connect it with the external legs. As the $SU(N)$ indices for two of the external $\phi$ legs in $M_n^{\text{NLSM}\oplus \phi^3}(\mathbb{I}_{n}|1,n,i)$ are adjacent to each other, it is clear that for the only vertex with three $\phi$, two of them must also have adjacent $SU(N)$ indices.

On the other hand, the RHS of Eq. (\ref{eqnlsmosex}) can  be seen as a sum of terms, each being a product of 1) momentum factor $s_{n+1,j}$ with $1<j<n$, 2) $2k+1$ NLSM semi-on-shell amplitudes, and 3) a constant $V_{2k+1}$ that depends on $j$. Diagramatically, each term can be seen  as a new odd-point vertex $V_{2k+1}$ connected by a number of $J(\alpha)$, multiplied by a momentum factor consistent with Eq. (\ref{eqchynlsm}); $V_{2k+1}$ is the only place where the Feynman rule on the RHS of Eq. (\ref{eqnlsmosex}) is different from that of NLSM. Naturally, this vertex is the one and only vertex with 3 $SU(\tilde{N})$ indices, and we reach observation 3. As Eq. (\ref{eqnlsmosex}) is the Ward identity and $V_{2k+1}^{\text{NLSM} \oplus \phi^3}$ is exactly where the current enters, its form is fixed by the structure, shown in Eq. (\ref{eq:nlsmc}), of the current $\mathcal{J}_\mu^a$ under the complete shift symmetry. For $k=1$, we have
\bea
V_3^{\text{NLSM}\oplus \phi^3}(1,2,3|1,3,2)  = - i \lambda,
\eea
recovering the self-interaction of $\phi$ given by Eq. (\ref{eq:phiself}).

Lastly, remember that the $SU(\tilde{N})$ flavor current needs to flow from  $V_{2k+1}^{\text{NLSM} \oplus \phi^3}$ to external legs, passing a number of vertices with 2 $SU(\tilde{N})$ indices. As the flavor current go through the semi-on-shell amplitudes of $\nlsm \oplus \phi^3$ which have exactly the same form as $J(\alpha)$ in NLSM, we arrive at observation 2. This also implies that the on-shell amplitude $M^{\nlsm \oplus \phi^3} (\alpha|i,j)$ with two $SU(\tilde{N})$ indices are exactly the same as the NLSM amplitude $M^\nlsm (\alpha)$, which is easy to see in the CHY construction of the extended theory as well \cite{Cachazo:2016njl}. As we know that $M^\nlsm (\alpha)$ is fixed by the soft constraint, namely the Adler's zero, of the NGBs, $M^{\nlsm \oplus \phi^3} (\alpha|i,j)$ must also be fixed by such constraints. This implies that even the biadjoint scalars must satisfy some form of shift symmetry when we consider the interactions involved in $M^{\nlsm \oplus \phi^3} (\alpha|i,j)$. However, such a symmetry is obviously broken by $V_{2k+1}^{\text{NLSM} \oplus \phi^3}$, as we know that $M_n^{\text{NLSM}\oplus \phi^3}(\mathbb{I}_{n}|1,n,i)$ does not vanish when we take the momentum of an external $\phi$ to zero. These observations offer hints on how to write down the Lagrangian of the extended theory, which we will see in Section \ref{sec:elag}.

\subsection{Extension of the special Galileon theory} 
\label{sec:sgale}

Similar to NLSM\ discussed in above, an extended theory was also identified in the single soft theorem of special Galileon in Ref. \cite{Cachazo:2016njl}:
\bea
M_n^\sgal = \frac{\tau^3}{4 \lambda \alpha^2} \sum_{a=2}^{n-2} \sum_{\substack{c=2\\ c\ne a}}^{n-1} \sum_{\substack{d=1\\d\ne a}}^{n-2} s_{an} s_{cn} s_{dn} M_{n-1}^{\sgal \oplus \text{NLSM}^2 \oplus \phi^3}(a,c,1|n-1,d,a) + \ordr (\tau^4),\label{eqsgalchy}
\eea
where leg $n$ is the soft leg. Again, we need to add the constant $4 \lambda \alpha^2$ to compensate for the mass dimension difference between the two sides of the equations. Here $\sgal \oplus \text{NLSM}^2 \oplus \phi^3$ denotes a mixed theory of four different kinds of scalars: the special Galileon $\pi$, the NGB $\Sigma^a$ that transforms under group $SU(N)$, another copy of NGB $\tilde{\Sigma}^{\tilde{a}}$ that transforms under $SU(\tilde{N})$, and a biadjoint scalar $\phi^{a \tilde{a}}$ that transforms under both $SU(N)$ and $SU(\tilde{N})$. We will use the shorthand notation $\pi+$ to denote the extended theory below. The CHY representation of such a theory is given by
\bea
M^{\pi+}_n (\alpha | \beta) = \oint d\mu_n \; \Big( \mathcal{C}(\alpha)\, (\pf\, \A_{\bar{\alpha}})^2 \Big) \Big( \mathcal{C}(\beta)\, (\pf\, \A_{\bar{\beta}})^2 \Big),\label{eq:chygm}
\eea
up to coupling constants that ensure the amplitude having the correct mass dimension. Details about the formula is reviewed in Appendix \ref{app:chyr}, the property of which will be useful in the following. We will denote the number of  labels in $\alpha / \beta$ as $\nf_{\alpha/\beta}$, and the total number of flavor labels $\nf \equiv \nf_{\alpha} +\nf_{\beta}$.

As before, in the flavor-ordered amplitude $M_{n-1}^{\pi+} (a,c,1|n-1,d,a)$, the left ordering is for $SU(N)$ and the right for $SU(\tilde{N})$. An index in both left and right orderings corresponds to a biadjoint scalar. If an index only shows in the left or right ordering, it belongs to an NGB. Indices that do not appear in the orderings come from Galileons. Considering the range of the sums on the RHS of Eq. (\ref{eqsgalchy}), we see that it involves amplitudes of different particle contents. For example, $M_{n-1}^{\pi+} (a,n-1,1|n-1,1,a)$ contains 3 $\phi$'s and $n-4$ Galileons, while $M_{n-1}^{\pi+} (a,c,1|n-1,1,a)$ with $c \ne n-1$ involves two $\phi$'s, one $\Sigma$, one $\tilde{\Sigma}$, and $n-5$ Galileons.

The soft theorem derived from the Ward identity, on the other hand, is given by Eq. (\ref{eqsgalsslf}):
\bea
M_n^\sgal &=&\tau^3 \left\{ \sum_{l} \frac{2}{\alpha^2}  p_{n, l^{1}} p_{n, l^{2}} p_{n, l^{3}} \prod_{i=1}^3 J ( \{ p_{l^i_j} \})\right.\nonumber\\
&&\left.- \sum_{l} \frac{1}{6 \alpha^4}  \sum_{\sigma \in S_5}   p_{n, l^{\sigma(1)}} p_{n, l^{\sigma(2)}} p_{n, l^{\sigma(3)}} \left[q_{l^{\sigma(4)}}^2 q_{l^{\sigma(5)}}^2 - p_{l^{\sigma(4)}, l^{\sigma(5)}}^2 \right] \prod_{i=1}^5 J ( \{ p_{l^i_j} \})\right\}\nonumber\\
&& + \ordr (\tau^4),\label{eq:sgaltsfw}
\eea
where we take the soft limit of $p_n$. We need to compare the above equation with Eq. (\ref{eqsgalchy}) to identify the Feynman rules of the extended theory. We  start with the lowest point amplitude, expressing the 4-pt amplitude of special Galileon using Eq. (\ref{eq:sgaltsfw}):
\bea
M_4^{\sgal} = \tau^3 \frac{1}{4 \alpha^2} s_{14} s_{24} s_{34} + \ordr (\tau^4),
\eea
while Eq. (\ref{eqsgalchy}) gives
\bea
M_4^{\sgal} = \frac{ \tau^3 }{4 \lambda \alpha^2} s_{14} s_{24} s_{34} M_3^{\pi +} (2,3,1|3,1,2) + \ordr (\tau^4),
\eea
in which $M_3^{\pi +} (2,3,1|3,1,2)$ only contains the cubic vertex of biadjoint scalars. Just like the case of the extension of NLSM, here we identify the cubic vertex to be
\bea
iM_3^{\pi +} (2,3,1|3,1,2) = V_3^{\pi+}  (2,3,1|3,1,2) = i \lambda.\label{eq:scalfr3}
\eea

Going to a higher point amplitude, on the other hand, is far more complicated here compared to the extension of NLSM. One reason is that as we discussed earlier, the RHS of Eq. (\ref{eqsgalchy}) involves several different kinds of amplitudes, each with different particle content. Another reason is that the momentum factors on the RHS of Eq. (\ref{eq:sgaltsfw}) are permutation invariant, while the momentum factors in Eq. (\ref{eqsgalchy}) are not. There can be a momentum factor of $(s_{in})^2 s_{jn}$ in Eq. (\ref{eqsgalchy}), but there are no such terms in Eq. (\ref{eq:sgaltsfw}). Apparently momentum conservation needs to be used here, but it is not straightforward to find the correct way to disentangle Eq. (\ref{eq:sgaltsfw}) and identify vertices.

However, there is also much similarity between the extensions of NLSM and special Galileon. For both of the tree level soft theorems derived from the Ward identity, namely Eqs. (\ref{eqnlsmosex}) and (\ref{eq:sgaltsfw}), we have some new vertices connected by semi-on-shell amplitudes of the original theory. In other words, the RHS of Eq. (\ref{eq:sgaltsfw}) can be seen as a sum of diagrams of purely special Galileon sub-diagrams connected by a new 3- or 5-pt vertex. In each of such diagrams, only one of the vertices has a Feynman rule different from that of special Galileon.

Another observation is that in the soft theorem given by the CHY formalism, new flavor indices arise in both the extensions of NLSM and special Galileon. As argued in the last section, there cannot be any vertex carrying only one flavor index of a certain group, so that there is a unique sub-diagram connecting the three external new flavor indices in the amplitudes involved in the NLSM single soft theorem.  In the extension of the NLSM, the vertex where the three new flavor flows intersect is precisely the new vertex given by the current $\mathcal{J}_\mu^a$. In the case of special Galileon, there are two groups of new flavor indices, then for each group there should exist a unique sub-diagram connecting the three external flavor indices. As we see from Eq. (\ref{eq:scalfr3}), the cubic vertex of biadjoint scalars exists in the amplitudes of special Galileon extension. This naturally leads us to conjecture that the two groups of external flavor flows in Eq. (\ref{eqsgalchy}) converge at the same vertex, which comes from the current under the enhanced shift symmetry -- more precisely, the $B$ term in Eq. (\ref{eqsgalwifi}).\footnote{Such a conjecture can be argued more rigorously using the CHY formula given by Eq. (\ref{eq:chygm}), which forbids interaction with an odd $\nf$.} Then all other vertices in the diagrams contain 0 or 2 indices for each of the two copies of $SU(N)$, and the Feynman rules for these vertices are exactly the same as the Feynman rules for special Galileon.

In summary, the above arguments lead to the following conjectures for vertices with at most 6 flavor indices, which are needed to construct the amplitudes $M^{\pi+} (a,c,1| n-1, d, a)$ on the RHS of Eq. (\ref{eqsgalchy}):
\begin{enumerate}
\item The total number of flavor indices of a vertex, $\nf$, must be even.
\item The allowed vertices with $\nf = 2$ are $\Sigma^2 \pi^2$ and $\tilde{\Sigma}^2 \pi^2$, the Feynman rules of which are the same as the one for the $\pi^4$ vertex.
\item The allowed vertices with $\nf = 4$ are $\phi^2 \pi^2$ and $\Sigma \tilde{\Sigma} \phi \pi$, the Feynman rules of which are the same as the one for the $\pi^4$ vertex.
\item The allowed vertices with $\nf = 6$ are $\phi^3$, $\phi^3 \pi^2$, $\Sigma \tilde{\Sigma} \phi^2 \pi$, and $\Sigma^2 \tilde{\Sigma}^2 \phi$.
\end{enumerate}
Just like the case in NLSM, the first three conjectures above ensure that a semi-on-shell $\pi+$ amplitude with $\nf_{\alpha/ \beta} = 2$ is exactly the same as the semi-on-shell amplitudes with $\nf_{\alpha/ \beta} = 0$, which can also be seen clearly in the CHY formula. The fourth observation implies that apart from $V^{\pi+}_3(1,2,3|1,2,3)$ given by Eq. (\ref{eq:scalfr3}), we have 3 more vertices to solve: $V^{\pi+}_5(1,2,3|1,2,3)$, $V^{\pi+}_5(1,2,3|1,2,4)$, and $V^{\pi+}_5(1,2,3|1,4,5)$.  Therefore, we can write Eq. (\ref{eqsgalchy}) in terms of the unknown vertices and semi-on-shell amplitudes, so that Eq. (\ref{eq:sgaltsfw}) can impose constraints on these unknown vertices. Moreover, the flavor-ordered objects satisfy some most general constraints, e.g. by definition they are invariant under a cyclic permutation of the indices. An additional constraint for flavor-ordered objects with 3 ordering indices is that we always get a minus sign when we reverse the ordering, e.g.
\bea
M(a,b,c|d,e,f)  =-M(b,a,c|d,e,f) = -M (a,b,c|e,d,f).\label{eq:sgeoce}
\eea
This is because we are considering the adjoint representation where the generator $\left(T^a \right)_{bc} = -if^{abc}$, so that the flavor factor with three indices, $\tr \{ T^a T^b T^c\} = - \tr\{ T^c T^b T^a \}$, is already fully anti-symmetric. Another constraint, which comes from the CHY formula Eq. (\ref{eq:chygm}), is that the flavor-ordered objects should stay the same when the left and right indices are exchanged, e.g.
\bea
M(a,b,c|d,e,f) = M(d,e,f|a,b,c).
\eea
The above relations also hold for flavor-ordered vertices and semi-on-shell amplitudes. The last observation is that in principle we can have a CP odd term like $\vep_{\mu \nu \rho \sigma} p_2^\mu p_3^\nu p_4^\rho p_5^\sigma$ in  $V^{\pi+}_5(1,2,3|1,4,5)$, but this is again excluded in the CP even CHY formula.

Unlike the case of the NLSM extension, even after imposing all constraints discussed above, we still cannot fix all of the unknown vertices using Eqs. (\ref{eqsgalchy}) and (\ref{eq:sgaltsfw}). The details of deriving the constraints on the vertices imposed by the Ward identity are shown in Appendix \ref{app:sgaled}. We are able to confirm the form of $V^{\pi+}_3(1,2,3|1,2,3)$ given by Eq. (\ref{eq:scalfr3}), and also fix the form of $V_5^{\pi+} (1,2,3|1,2,3)$: 
\bea
 V_5^{\pi+} (1,2,3|1,2,3) &=& -\frac{i\lambda}{  \alpha^2} \left[ p_4^2 p_5^2 - p_{4,5}^2 \right]
\eea
On the other hand, we can identify solutions for $V^{\pi+}_5(1,2,3|1,2,4)$ and $V^{\pi+}_5(1,2,3|1,4,5)$, but the solutions are parameterized by 8 free coefficients. As the RHS of Eq. (\ref{eqsgalchy}) contains a specific combination of  amplitudes instead of individual ones, the unfixed contributions cancel each other in the sum and thus does not enter the soft limit of $M_n^\sgal$. For example, one solution is
\bea
V_5^{\pi+} (1,2,3|1,2,4) &=& -\frac{i\lambda}{ \alpha^2} \left(  p_{3,5} p_{4,5} - p_{3,4} p_5^2 \right),\label{eq:crctv1}\\
V_5^{\pi+} (1,2,3|1,4,5) &=& -\frac{i\lambda }{  \alpha^2}  \left( p_{1,3} p_{2,4} - p_{1,4} p_{2,3} \right),\label{eq:crctv2}
\eea
while another solution is
\bea
V_5^{\pi+} (1,2,3|1,2,4) &=& -\frac{i\lambda}{ \alpha^2} \left( p_{3,5}p_{4,5} - p_{3,4} p_5^2 - p_{1,4} p_{2,3} - p_{1,3} p_{2,4} \right),\non\\
V_5^{\pi+} (1,2,3|1,4,5) &=& -\frac{2i\lambda }{  \alpha^2}  \left( p_{1,3} p_{2,4} - p_{1,4} p_{2,3} \right)
\eea
The difference of these two solutions is cancelled out in the sum of amplitudes on the RHS of Eq. (\ref{eqsgalchy}). However, the different choices of vertices are not equivalent: the above two solutions apparently give different results for the 5-pt amplitude $M^{\pi+}_5 (1,2,5| 3,4,5)$. Therefore, unlike in NLSM and its extension, where different parameterizations of the Lagrangian can be equivalent and lead to the same on-shell amplitudes (an example of the NLSM extension is given in Ref. \cite{Low:2018acv}), the difference here cannot be explained by unfixed parameterization of the NGB fields involved. Therefore, we need even more constraints to fix the Feynman vertices. It turns out that the constraints that we need come from the soft behavior of the extended theory. If we impose that for the amplitudes in the extended theory, the leading single soft behavior for $\gs$ or $\gst$ is at $\ordr (\tau)$, and the leading single soft behavior of the Galileon $\pi$ is at $\ordr (\tau^2)$, one can completely fix the vertices to be given by Eqs. (\ref{eq:crctv1}) and (\ref{eq:crctv2}). Such soft behaviors can also be easily proved using the CHY formula, therefore can be seen as another kind of constraint imposed by Eq. (\ref{eq:chygm}) that we need to consider.

In summary, just like the extension of NLSM, we have found the unique forms of local interactions in the extension of special Galileon. For both cases, apart from the results of the Ward identity, we have utilized two kinds of additional constraints. The first kind is about the ordering of the vertices, such as Eq. (\ref{eq:sgeoce}). The flavor ordering happens because there is some unbroken linearly realized symmetry, thus the constraining the vertices using its ordering property is equivalent to imposing the unbroken symmetry. The second kind of constraint is about the soft behavior of the amplitude, and we know that such behavior is protected by some nonlinearly realized shift symmetry. Therefore, by fixing the Feynman rules of the extended theory, we have recognized the existence of both linear and nonlinear symmetries in the extended theories.

\subsection{The Lagrangian of the extended theories}
\label{sec:elag}

Given the Feynman rules discussed in the previous sections, we now write down a Lagrangian that generates the vertices we need. Attempts for $\nlsm \oplus \phi^3$ have been made before in Ref. \cite{Cheung:2016prv}, which presented the Lagrangian in terms of alternative variables  obscuring the Bose symmetry. Here we try to write down the Lagrangian using traditional variables, so that we can easily discover the symmetries implied. As the theory $\nlsm \oplus \phi^3$ is clearly part of the theory $\pi+$, we  directly write down the Lagrangian for the full theory $\pi+$, which contains flavorless $\pi$, $\gs$ in the adjoint of $SU(N)$, $\gst$ in the adjoint of $SU(\tilde{N})$ and $\phi$ in the adjoint of $SU(N) \times SU(\tilde{N})$. We can write the Lagrangian for the full theory as follows:
\bea
\lag^{\pi+} = \lag^\gs + \lag^\pi + \lag^\phi,
\eea
where we call $\lag^\gs$, $\lag^\pi$ and $\lag^\phi$ as the ``NG sector'', the ``Galileon sector'' and the ``cubic sector'', respectively. When we set $\pi = 0$ and $\gst = 0$, $\lag^{\pi^+}$ is reduced to the Lagrangian of $\nlsm \oplus \phi^3$.

The NG sector is given by
\bea
\lag^\gs &=& \frac{f^2}{8 (N^2-1) } \tr \left\{\partial_\mu \left[U \left(\mPsi^+ \right) \right]^{-1} \partial^\mu U \left(\mPsi^+ \right) + \left( \mgs \leftrightarrow \mgst \right)\right\} - \frac{1}{2} \partial_\mu \phi^{a\ta} \partial^\mu \phi^{a\ta},\label{eq:laggs}
\eea
where $U(\underline{x})$ is a function of matrix $\underline{x}$ satisfying $[U(\underline{x})]^{-1} = U(-\underline{x})$ as well as $U(0) = 1$, $\mPsi^\pm \equiv \mphi \pm \mgs$, and
\bea
 \left(\mgs \right)_{ab, \ta \tb} &\equiv&  \frac{2i}{f}\  \gs^c (T^c)_{ab} \delta_{\ta \tb}, \\
  \left(\mgst \right)_{ab, \ta \tb} &\equiv&  \frac{2i}{f} \  \gst^\tc \delta_{ab} (T^\tc)_{\ta \tb},\\
    \left(\mphi \right)_{ab, \ta \tb} &\equiv&  \frac{2i}{f} \  \sqrt{N^2-1} \phi^{c \tc} (T^c)_{ab} (T^\tc)_{\ta \tb}.
\eea
As reviewed in Appendix \ref{app:nlsmpara}, when we set $\phi = 0 $, $\lag^\Sigma$ will give the 2-derivative NLSM Lagrangian of $\gs^a$ and $\gst^\ta$ under a general parameterization. For the exponential parameterization corresponding to the infrared construction of NLSM presented in Section \ref{sec:nlsm}, we have $U(x) = e^x$ . If we recognize $\delta_{ab}/\sqrt{N^2 - 1}$ as the properly normalized $U(N)$ generator that corresponds to the $U(1)$ subgroup, we see that  Eq. (\ref{eq:laggs}) contains actually the 2-derivative Lagrangian of two copies of NLSM, both coming from a coset like $H \times H/ H$ reviewed in Appendix \ref{app:diagnlsm}: one copy has $H = SU(N) \times U(\tilde{N})$, while the other has $H = U(N) \times SU(\tilde{N})$.  We already know that the NLSM Lagrangian satisfies the shift symmetry, and from the observation 2 discussed in Section \ref{sec:nlsme}, we know that the even-pt interactions with two $\phi$ in $\nlsm \oplus \phi^3$ also need to satisfy such a symmetry. We see clearly how this is realized in $\lag^\gs$: $\phi$ becomes the NGB of $SU(N) \times U(\tilde{N})$ NLSM. Such a construction ensures that for a $v$-pt vertex with $\nf_\alpha = v$ and $\nf_\beta = 2$, the Feynman rule is exactly the same as a vertex with $\nf_\alpha = v$ and $\nf_\beta = 0$. This means that we have all the even-pt vertices we need for $\nlsm \oplus \phi^3$.

The Galileon sector is given by
\bea
\lag^\pi &=& \frac{1}{2} \partial_\mu \pi \partial^\mu \pi- \frac{1}{24 \alpha^2} \tr \left\{ \mPhi \lag_3^\TD (\mPhi) \right\},
\eea
where
\bea
\mPhi \equiv \mpi + \mgs+\mgst + \mphi,\qquad \left(\mpi\right)_{ab,\ta \tb} \equiv  \frac{2i}{f}  \frac{1}{\sqrt{N^2-1}} \pi \delta_{ab} \delta_{\ta \tb}.
\eea
Here we have promoted $\pi$ to $\mPhi$ in the interaction term in $\lag_\sgal$. This is the single trace multi-field Galileon vertex discussed in Ref. \cite{Cheung:2016drk}. It generates the pure special Galileon interaction, and also ensures that a vertex with $\nf_{\alpha/ \beta} = 2$ is the same as a vertex with $\nf_{\alpha/ \beta} = 0$. Therefore, we have all the even-pt vertices we need for $\pi+$.

\begin{figure}[htbp]
\centering
\includegraphics[height=3.8cm]{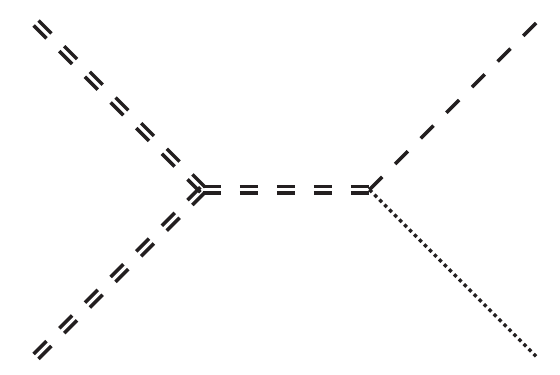}
\caption{A possible contribution to $M^{\pi+}(1,2,3|1,2,4)$, where $\gs$, $\gst$ and $\phi$ are represented by single dashed lines, dotted lines and double dashed lines, respectively.\label{fig:n4f6p}}
\end{figure}

One may be tempted to write $\lag^\gs$ as below:
\bea
\frac{f^2}{8 (N^2-1) } \tr \left\{\partial_\mu \left[ U \left(\mPhi  \right) \right]^{-1} \partial^\mu U \left(\mPhi  \right) \right \},\label{eq:uunlsm}
\eea
which utilizes $\mPhi$ just like in $\lag^\pi$.
Eq. (\ref{eq:uunlsm}) is the Lagrangian for $U(N)\times U(\tilde{N})$ NLSM, while the Galileon $\pi$ corresponds to the $U(1)\times U(\tilde{1})$ subgroup and completely decouples. The problem of such a term is that it generates non-zero vertices of $\phi^2 \gs^{2k-1} \gst^{2n-2k-1}$, which should be forbidden in the CHY construction given by Eq. (\ref{eq:chygm}). For example, there is a non-vanishing vertex $\phi^2 \gs \gst$, while the amplitude $M_4^{\pi+} (1,2,3|1,2,4)=0$. In principle such a vertex can be cancelled by a diagram shown in Fig. \ref{fig:n4f6p}, where we need to introduce new vertices like $\phi \gs \gst$. For now we do not see how such a vertex can arise, thus we are satisfied with $\lag^\gs$ given by Eq. (\ref{eq:laggs}).

The way to write down the cubic sector, which contains all the odd-pt vertices that we need, is less intuitive. Let us first consider the odd-pt vertices we need in $\nlsm \oplus \phi^3$. It is known that when we take an NGB in $M^{\nlsm \oplus \phi^3}_{2n+1} (\mathbb{I}_{2n+1}|1, 2n+1, i)$ to be soft, the amplitude still vanishes \cite{Cachazo:2016njl}. In other words, Adler's zero condition is still satisfied for the NGB's in such an amplitude. As reviewed in Appendix \ref{app:diagnlsm}, this means that we need to construct objects using $U(\mgs)$, adding $\phi$ to it to break the shift symmetry. As the coset associated with the symmetry breaking that results in the NGBs, $H \times H/H$ with $H= SU(N)$, is a symmetric coset, the interaction should also be invariant under $\mgs \to -\mgs$. The simplest term we can write down that satisfies all of the above is
\bea
i\tr \{ U(\mPsi^+ ) U(\mPsi^-) \} + \text{h.c.},\label{eq:lcsft}
\eea
where $\text{h.c.}$ means hermitian conjugation. Eq. (\ref{eq:lcsft}) already gives the correct coefficients of the odd-pt vertices with 3 $\phi$ in the exponential parameterization of $\nlsm \oplus \phi^3$, up to an overall real constant factor. However, such a term fails to give the correct vertices in Cayley parameterization discussed in Appendix \ref{app:nlsmpara}. The term that works for both the exponential and Cayley parameterizations turns out to be
\bea
i\tr \{ [U(\mPsi^+) - U(\mgs)] [ U(\mPsi^-) - U(-\mgs)] \} + \text{h.c}.
\eea
In the exponential parameterization, the above includes the interaction terms
\bea
&& \frac{i}{2} \sum_{k=2}^{\infty} \sum_{j=2}^{2k-1} \frac{(-4)^{k}  \lambda}{ (2k+1)! f^{2k-2}  } \left[\left( \begin{array}{c}
2k\\
j-1
\end{array} \right)  (-1)^{j-1} - 1 \right] \non\\
&\times& \tr \left\{ \phi^{a\ta} T^a \phi^{b\tb} T^b \left( \gs^d T^d  \right)^{j-2} \phi^{c \tc} T^c \left( \gs^e T^e \right)^{2k-j} \right\} f^{\ta \tb \tc}  ,
\eea
up to an overall constant, which give the Feynman vertices in Eq. (\ref{eqmixedfr}). On the other hand, we only have a limited number of odd-pt vertices appearing in the single soft theorem of special Galileon, so we can just list all the corresponding operators in the Lagrangian.
In the end, we can write $\lag^\phi$ as
\bea
\lag^\phi &= &  \frac{5}{6} \lambda  \phi^{a \ta} \phi^{b \tb} \phi^{c \tc} f^{abc} f^{\ta \tb \tc} +  \frac{\lambda}{12 \alpha^2} \phi^{a \ta} \phi^{b \tb} \phi^{c \tc} f^{abc} f^{\ta \tb \tc} \lag_2^\TD (\pi) \non\\
&&-\frac{\lambda}{2 \alpha^2} \phi^{a \ta} \phi^{b \tb}  f^{abc} f^{\ta \tb \tc}  \partial_\mu \gs^c \partial_\nu \gst^\tc T_2^{\mu \nu} (\pi)  + \frac{\lambda}{2 \alpha^2} \phi^{a \ta}   f^{abc} f^{\ta \tb \tc} \partial_\mu \gs^b \partial_\nu \gs^c \partial^\mu \gst^\tb \partial^\nu \gst^\tc \non\\
&&+  \left[ \frac{6if^3 \lambda}{ (N^2-1)^{3/2}} \tr \left\{[U(\mPsi^+) - U(\mgs)] [ U(\mPsi^-) - U(-\mgs)] + \left( \mgs \leftrightarrow \mgst \right)  \right\}  + \text{h.c.}\right].
\eea
Notice that when we exchange $\mgs$ with $\mgst$ in the above, the component $\mgs$ contained in $\mPsi^\pm \equiv \mphi \pm \mgs$ needs to be exchanged with $\mgst$ as well.

The Lagrangian $\lag^{\pi+}$ that we write down also introduces lots of new interaction vertices that do not enter the amplitudes involved in the single soft theorems of NLSM and special Galileon amplitudes. Nevertheless, one can already see that most of the extra terms do not enter the CHY formula given by Eq. (\ref{eq:chygm}): they contain such a high power of momenta that they can only act as higher order corrections to Eq. (\ref{eq:chygm}). More specifically, Eq. (\ref{eq:chygm}) contains $m=2(3+n-\nf)$ powers of momenta, the details about which is discussed in Appendix \ref{app:chyr}. Let us define the ``weight'' $w \equiv (m/2)+\nf - n$, so that $w =3$ is a constant in Eq. (\ref{eq:chygm}).\footnote{The existence of such a ``weight'' is hinting flavor-kinematics duality, as we are putting momentum and flavor on equal footing.} Such a definition for the weight can also be applied for vertices and diagrams. It turns out that the weight of a diagram cannot be decreased by adding new vertices to it. Then a term in $\lag^\gs$ like $\phi^4 \gs^2$ with a weight of $w = 5$ can never enter Eq. (\ref{eq:chygm}): it has too many powers of momenta.

\begin{figure}[htbp]
\centering
\includegraphics[height=3cm]{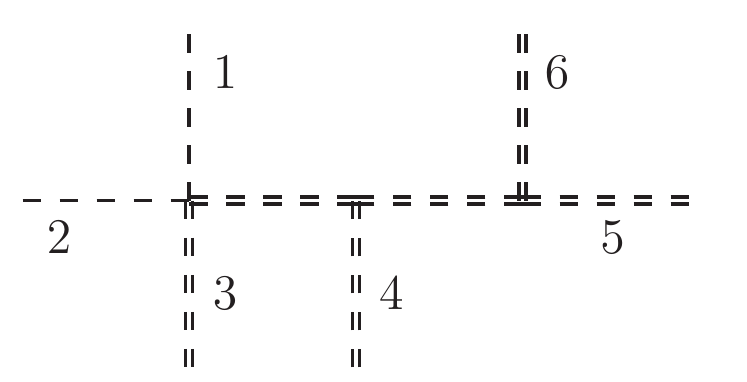}
\caption{A diagram in $M_6^{\pi +} (\mathbb{I}_6 | 3,4,5,6)$, where $\gs$ and $\phi$ are represented by single dashed lines and double dashed lines, respectively.\label{fig:e1636}}
\end{figure}

For example, let us consider the amplitude $M_6^{\pi +} (\mathbb{I}_6 | 3,4,5,6)$ calculated in Ref. \cite{Cachazo:2016njl}, which has $m=-2$, $\nf = 10$ and $n=6$, thus $w = 3$. An example of the diagrams contributed to such an amplitude is given by Fig. \ref{fig:e1636}, which involves $\phi^3$ and $\phi^2 \gs^2$ vertices that also appear in the single soft theorem of NLSM. The $\phi^4 \gs^2$ vertex in $\lag_\gs$ with $m=2$ is clearly a higher order correction to such an amplitude, therefore do not enter the CHY formula.

It can be checked that the only vertex that 1) appears in $\lag^{\pi+}$, 2) does not appear in the soft limit of NLSM or special Galileon, and 3) has a power of momenta low enough to enter Eq. (\ref{eq:chygm}) is the $\gs^2 \gst^2$ vertex in $\lag^{\pi}$. Such a vertex needs to exist to guarantee that $M^{\pi+}_n (i,j| k,l) = M^{\sgal}_n$.

\section{Conclusion and discussions}
\label{sec:conclusion}
In this work we presented the subleading single soft limit of several scalar EFTs. In particular, the on-shell amplitudes of NLSM, DBI and special Galileon possess non-trivial single soft limits, which are known to be related to the underlying shift symmetry of the theories. We performed a systematic analysis of these theories using quantum field-theoretic methods, by deriving the Ward identities associated with the shift symmetry. We showed that the (enhanced) Adler's zero of these theories is a direct consequence of the Ward identities. Moreover, we went one step beyond the (enhanced) Adler's zero and derived the subleading single soft limit of these theories.

By reviewing the Ward identity and subleading single soft theorem of NLSM, we showed that the key ingredient in our method is the complete form of the nonlinear shift. For the theory of DBI scalars, we derived the Ward identities for both the constant shift and the enhanced shift symmetry, and showed explicitly that only in the latter is the enhanced Adler's zero explicit. At tree level, we obtained novel Berends-Giele recursion relations, and expressed the soft limit of an on-shell amplitude as semi-on-shell sub-amplitudes connected by a new vertex proportional to two powers of the soft momentum. A similar treatment of the ordinary Galileon theory led to the conclusion that the Ward identity of the constant shift is enough to protect the enhanced soft behavior, so that the soft behavior of Galileon theory is not exceptional. On the other hand, the Ward identity for the full nonlinear shift of the special Galileon theory is indeed necessary to ensure its exceptional soft behavior. Tree level results where the enhanced soft behavior becomes explicit were also presented for both the ordinary and the special Galileon theory.

A well-established result proved using the amplitude method is that there are no exceptional scalar theories with the single soft limit of $\ordr (\tau^s)$, where $s \ge 4$ \cite{Cheung:2016drk}. It would be interesting to see how such a constraint can emerge at the Lagrangian level. From the story of ordinary Galileon, we may conjecture that a scalar theory is only exceptional if its full soft behavior is protected by a nonlinear shift symmetry beyond the simple constant shift, i.e. if its $\ordr (\tau^s)$ soft behavior can only become transparent when we derive the Ward identity corresponding to an enhanced shift symmetry. We leave such a conjecture for future examination.

Our tree level results of the subleading single soft theorems of NLSM and special Galileon provided us with a handle to study the properties of the emergent extended theories discovered using the CHY formalism. We reviewed how the Feynman vertices of the extension of NLSM were identified. For the extension of the special Galileon theory, the identification of the Feynman vertices is far more complicated, and it is useful to apply the lessons learned from studying the extension of NLSM so that we could take an educated guess of the flavor structure in the relevant amplitudes. Additional input of the CHY formula is also needed to completely fix the relevant vertices. The bottom line is that for both the extensions of NLSM and the special Galileon theory, we are able to find unique solutions\footnote{For the extension of NLSM, the form of the local interactions is unique only when we fix the parameterization of the NLSM Lagrangian.} of the relevant  interactions using the ordering property as well as the soft behavior of the amplitudes, which imply the linear and non-linear symmetry of the underlying theory, respectively.

We then presented the full Lagrangian $\lag^{\pi+} = \lag^\gs + \lag^\pi + \lag^\phi$ of the extended theory of special Galileon. The form of $\lag^\gs$ and $\lag^\pi$ is hinting that (enhanced) shift symmetry is governing all kinds of scalar fields involved, even including the biadjoint scalars. On the other hand, whether the Lagrangian $\lag^{\pi+}$ is complete or not is still an open question. From the perspective of symmetry, there may be better ways to write the  $\ordr (p^4)$ terms in $\lag^\phi$. We have written the $\ordr (p^0)$ terms in a way that manifests the underlying shift symmetry of the NGBs, using objects like $U(\mgs )$ and $U(\mPsi^\pm)$ as building blocks. As we know that $M^{\pi+}_{n-1} (a,c,1|n-1,d,a)$ also displays enhanced soft behavior when a Galileon is taken to be soft, we should also be able to write the $\ordr(p^4)$ terms in a way that shows the underlying symmetry that is protecting such a behavior. From the perspective of the CHY representation, it remains to be seen whether $\lag^{\pi+}$ provides all the necessary vertices that enter the existing CHY formulae. Apart from the single trace formulae presented in Ref. \cite{Cachazo:2016njl}, the latest development \cite{Mizera:2018jbh} also shows the CHY formulae for multi-trace terms in $\nlsm \oplus \phi^3$, with many examples for the low-pt amplitudes in the theory. It will be interesting to compare $\lag^{\pi+}$ with both the multi-trace results and those of the Z-theory \cite{Carrasco:2016ygv,Mizera:2018jbh}.

A more general question that is related to the discussion above is what the guiding principle is when we write down $\lag^{\pi+}$. Recently a much discussed topic is using operators to transform the amplitudes of one theory to another \cite{Cheung:2017ems,Zhou:2018wvn,Bollmann:2018edb}, which manifests color/flavor-kinematic duality. It is not clear how one should realize it at the level of Lagrangian. Promoting $\pi$ to $\mPhi$ in $\lag^\pi$ is a clear example, but it is still an open question how the exact form of $\ordr (p^0)$ term in $\lag^\phi$ is generalized from the NLSM Lagrangian and the cubic interaction of $\phi$.

Another pressing question is why different kinds of shift symmetry are present within the same theory $\pi +$. Is there a way to unify the symmetries? One thing that is clear is that the emergent mixed theory has its origin in flavor-kinematics duality, similar to the well-known color-kinematics duality seen in amplitudes of gauge theory and gravity \cite{Bern:2008qj}. More work needs to be done to see how such duality arises in the Lagrangian. For example, the representation for $\nlsm \oplus \phi^3$ as well as special Galileon in Refs. \cite{Cheung:2016prv,Cheung:2017yef,Mizera:2018jbh} makes the duality manifest using alternative variables, and it would be helpful if such a representation can also be found for the full extended theory $\pi+$.

There are also more scalar theories with non-trivial shift symmetries \cite{Griffin:2014bta}, and we do not know whether they have extensions or not. One can certainly derive Ward identities corresponding to the shift symmetries, but it  remains to be seen whether we can interpret the results as the amplitudes of some extended theories. Moreover, in Ref. \cite{Cachazo:2016njl} there are other extended theories that arise from amplitudes of fermionic and vector fields, both related to the Born-Infeld action. Our method of deriving Ward identities may be generalized beyond scalar EFTs to study these extended theories as well. One apparent starting point is to single out the scalar sector of the extended DBI and compare with the Ward identity of DBI presented in this work.

Apart from the extended theories, another direction worth exploring is the quantum aspects of the soft theorems. Loop corrections may affect the relations between the scalar EFTs and their extensions. Also, we have always assumed that the classical symmetry survives quantization in our treatment. Although this can be set as a constraint for the EFTs, how it is realized is a nontrivial matter. It is well-known that Wess-Zumino-Witten terms \cite{Wess:1971yu,Witten:1983tw} arise as a consequence of anomalies in NLSM. It would be interesting to see how they can be generated using the infrared construction of NLSM, without  recourse to current algebra. It is also unknown how to recover the higher derivative corrections of NLSM using pure amplitude methods. Recent works \cite{Elvang:2018dco,Rodina:2018pcb,Mizera:2018jbh} present interesting single trace results, while NGB interactions at four derivative level clearly contains double trace contributions. On the other hand, with the tools at hand we can study the soft limit of the NGB interactions beyond the two derivative Lagrangian, and the four derivative terms have important phenomenological applications for the study of the Higgs boson in new physics models \cite{Liu:2018vel,Liu:2018qtb}.  Scalar EFTs also appear in a wide range of models of cosmology, and the shift symmetry in these models leads to  important observable effects \cite{Finelli:2018upr}. Our Ward identities corresponding to the shift symmetry may provide new insights for this line of research.

\begin{acknowledgments}

The author would like to thank Ian Low for suggesting the problem, as well as many helpful discussions and comments on the manuscript. He also thanks John Joseph Carrasco, Bo Feng, Marios Hadjiantonis, Song He, Callum R.T. Jones, Shruti Paranjape, Laurentiu Rodina and Chia-Hsien Shen for useful discussions.

\end{acknowledgments}

\begin{appendix}

\section{Useful Properties of the Galileon Lagrangian}
\label{app:gall}

The interaction terms in the Lagrangian of the Galileon theory have less than two derivatives per field, and can be written as \cite{Nicolis:2008in}
\begin{align}
\lag^{(1)}_{n+1} = \partial_\mu \pi \partial_\nu \pi T^{\mu \nu}_n.\label{eqlag1}
\end{align}
where  the symmetric tensor $T^{\mu \nu}$ is given by Eq. (\ref{eq:galltdd}):
\begin{align}
T^{\mu \nu}_n \equiv \sum_{\sigma \in S_n} \epsilon (\sigma) g^{\mu}_{ \mu_1} g^{\nu \mu_{\sigma (1)}} \prod_{i=2}^n \Pi^{\mu_{\sigma (i)}}_{\mu_i}.
\end{align}
Because of the anti-symmetry given by $\epsilon (\sigma)$, $\partial_\mu T^{\mu \nu}_n = 0$, and in $d$ dimensions $T_n^{\mu \nu} = 0$ when $n > d$. $T_n$ satisfies a useful recursion relation:
\begin{align}
T^{\mu \nu}_n = g^{\mu \nu} \lag_{n-1}^\TD - (n-1) \Pi^\mu_{\rho} T^{\nu \rho}_{n-1},\label{eqtrr}
\end{align}
where
\begin{align}
\lag_n^{\TD} \equiv \Pi_{\mu \nu} T^{\mu \nu}_n =  \sum_{\sigma \in S_n} \epsilon (\sigma)  \prod_{i=1}^n \Pi^{\mu_{\sigma (i)}}_{\mu_i}.
\end{align}
Notice that $\lag^\TD_n$ can be written as a total derivative because of the anti-symmetry:
\begin{align}
\lag_n^{\TD} &=  \partial_{\mu } \left( \partial_\nu \pi T^{\mu \nu}_n \right) = \partial_{\mu} \partial_{\nu} \left( \pi T^{\mu \nu}_n \right).\label{eqpltd}
\end{align}

Under the Galilean transformation $\pi \to \pi + \vep + a_\mu x^\mu$, $ \lag^{(1)}_{n+1}$ transforms as
\bea
 \lag^{(1)}_{n+1}  \to \lag^{(1)}_{n+1} + 2 a_\mu \partial_\nu \pi T^{\mu \nu}_n =\lag^{(1)}_{n+1} + 2 a_\mu \partial_\nu \left( \pi T^{\mu \nu}_n  \right).\label{eql1gs}
\eea
Namely, the variation of the Lagrangian is a total derivative, so that the theory is invariant under the shift. An equivalent way to write $\lag^{(1)}_{n+1}$, which is more convenient for our purpose, is \cite{Deffayet:2011gz,Deffayet:2013lga}
\bea
\lag_{n+1}^{(2)} = \lag^{(1)}_{n+1} -  \partial_{\mu} \left[ \pi \partial_{\nu} \pi T^{\mu \nu}_n \right] =  -\pi \lag_n^{\TD}.
\eea
It is easy to check that because of the two total derivatives of $\lag_n^\TD$, under the shift the variation of $\lag_{n+1}^{(2)}$ can still be written as a total derivative:
\bea
\delta \lag_{n+1}^{(2)} \to \delta \lag_{n+1}^{(2)} - (\vep + a_\mu x^\mu) \lag_n^\TD,
\eea
where
\bea
x^\mu \lag_n^\TD = x^\mu \partial_\nu \partial_\rho (\pi T^{\nu \rho}_n) =  \partial_\nu \left[  x^\mu \partial_\rho (\pi T^{\nu \rho}_n) \right] - \partial_\rho (\pi T^{\mu \rho}_n).\label{eqggalxltd}
\eea
Therefore, the Lagrangian in Eq. (\ref{eq:ggallag}) transforms as
\bea
\lag_\gal \to \lag_\gal - \frac{1}{2}  \sum_{n=1}^d c_{n+1} \partial_{\mu} \left[ (\vep + a\cdot x) \partial_{\nu} \left( \pi T^{\mu \nu}_n \right) - a_\nu \pi T^{\mu \nu} \right],
\eea
and the Galilean symmetry is preserved.

Using the properties discussed above, one can prove that the RHS of Eq. (\ref{eqchmf}) can indeed be written as a total derivative. We know that
\bea
 x^{\{\mu } x^{ \nu \}} \lag_n^\TD =   \partial_\rho \left[ x^{\{\mu } x^{ \nu \}} \overleftrightarrow{\partial_\lambda} \pi T^{ \rho \lambda}_n \right] + 2 \pi T_{n }^{\{\mu \nu\} } ,
\eea
where the operator $\overleftrightarrow{\partial}$ is defined as $a\overleftrightarrow{\partial} b \equiv a \partial b - b \partial a$. From Eq. (\ref{eqtrr}) we know that
\bea
\pi T_n^{\{ \mu \nu\}} &=&  \frac{n(n-1)}{2} \partial^{\{\mu} \pi \partial^{\nu\}} \pi \lag^\TD_{n-2} - (n-1) \partial^{\{\mu} \left(\pi \partial^{\nu \}} \pi \lag_{n-2}^\TD \right) \non\\
&&+ \frac{(n-1)(n-2)}{2} \partial_\rho \partial_\lambda \left( \pi \partial^{\{ \mu}\pi  \partial^{\nu \}} \pi T_{n-2}^{\rho \lambda} \right) \non\\
&&- (n-1)(n-2) \partial_\rho \left(\partial_\lambda \pi \partial^{\{ \mu}\pi  \partial^{\nu \}} \pi T_{n-2}^{\rho \lambda} \right).\label{eqsgalkeyr}
\eea
As we work in 4 dimensions, the LHS of Eq. (\ref{eqsgalkeyr}) vanishes when $n>4$. Then
\bea
&&\left( \alpha^2 x^{\{ \mu} x^{ \nu \} } + \partial^{\{\mu } \pi \partial^{\nu \} } \pi \right) \partial_\rho J^\rho_\sgal \non\\
&=& \alpha^2 \partial_\rho \left[ x^{\{ \mu} x^{ \nu \} } \overleftrightarrow{\partial_\lambda} \pi T_1^{\rho \lambda} \right] - \frac{1}{6}  \partial_\rho \left[ x^{\{ \mu} x^{ \nu \} } \overleftrightarrow{\partial_\lambda} \pi T_3^{\rho \lambda} \right] \non\\
&& + \frac{2}{3} \partial^{\{ \mu } \left[ \pi \partial^{\nu \}} \pi \lag^{\TD}_1 \right] - \frac{1}{15 \alpha^2 } \partial^{\{ \mu} \left[ \pi \partial^{\nu \} } \pi \lag^{\TD}_3 \right]\non\\
&& + \frac{2}{3} \partial_{\rho} \left[ \partial^{\{ \mu } \pi  \partial^{\nu \}}  \pi \partial_{\lambda } \pi  T^{\rho \lambda}_1 \right] - \frac{1}{5 \alpha^2} \partial_\rho \left[ \partial^{\{ \mu } \pi  \partial^{\nu \}}  \pi \partial_{\lambda } \pi  T^{\rho \lambda}_3 \right]\non\\
&& - \frac{1}{3} \partial_{\rho}  \partial_{\lambda }\left[\pi \partial^{\{ \mu } \pi  \partial^{\nu \}}   \pi  T^{\rho \lambda}_1 \right] + \frac{1}{10 \alpha^2} \partial_\rho  \partial_{\lambda } \left[ \pi \partial^{\{ \mu } \pi  \partial^{\nu \}}   \pi  T^{\rho \lambda}_3 \right].
\eea

\section{Various aspects of NLSM}
\label{app:vanlsm}
In this section we review the flavor ordering of NLSM, the construction of the NLSM Lagrangian with the coset of $H \times H/H$, and different parameterizations of such a Lagrangian.

\subsection{Flavor ordering}
\label{app:flavo}
The vertices from the general two derivative Lagrangian of NLSM, given by Eq. (\ref{eq:nlsmlag}), can be written as \cite{Low:2017mlh,Low:2018acv}
\bea
&&\left[V^{\nlsm}_{2n} (p_1, \cdots, p_{2n}) \right]^{a_1 a_2 \cdots a_{2n}} \non\\
&=&  \sum_{\sigma \in S_{2n-1}}\tr \{ X^{a_{\sigma(1)}} X^{a_{\sigma(2)}} \cdots X^{a_{\sigma(2n-1)}} X^{a_{2n}}\}\non\\
&&\times i \frac{(-1)^{n}}{(2n)!} \left(\frac{4}{f^2}\right)^{n-1} \sum_{k=1}^{2n-1} (-1)^{k-1} \left( \begin{array}{c}
2n-2\\
k-1
\end{array}\right) \sum_{i=1}^{2n} (p_{\sigma (i)} \cdot p_{\sigma (i+k)} ),\label{eqnlsmnpve}
\eea
where $\sigma$ is a permutation of $\{1,2, \cdots, 2n-1\}$, $\sigma (2n) \equiv 2n$ and $\sigma (2n+k) \equiv \sigma (k)$. Generators $T^i$ in Eq. (\ref{eq:nlsmlag}) are of the unbroken group $H$, while the single trace flavor factor in Eq. (\ref{eqnlsmnpve}) contains broken generators $X^a$ associated with the coset $G/H$.  To arrive at Eq. (\ref{eqnlsmnpve}), we need to identify that $T^i_{ab} = -if^{iab}$ \cite{Low:2014nga}, where $f^{iab} = -i\tr \{ T^i [X^a, X^b] \}$ is the structure constant of $G$. The coset also need to be symmetric, so that $[X^a, X^b] = if^{iab} T^i$.

It can be proved that the flavor ordered amplitudes can also be defined. However, in general the relation between the flavor ordered vertices and the amplitudes is non-trivial and $G$ dependent. The case becomes simpler when two  traces of $X^a$ can be merged into a single trace, so that we can directly use the flavor-ordered Feynman vertices to construct diagrams. The well-studied example is for the coset of $SU(N)_L \times SU(N)_R / SU(N)_V$ \cite{Kampf:2013vha}, where the space expanded by  $X^a$ is isomorphic to the space expanded by $T^i$, so that the completeness relation of the adjoint of $SU(N)$ group can be applied to merge traces:
\bea
\tr \{T^i \underline{A} \} \tr \{T^i \underline{B}\} = \tr\{\underline{A}\, \underline{B}\} - \frac{1}{N} \tr\{\underline{A}\} \tr\{ \underline{B} \},
\eea
where $\underline{A}$ and $\underline{B}$ are products of generators. In an amplitude the disconnected term on the RHS of the above equation will drop out. This can be understood as the consequence of $U(1)$ decoupling in the $U(N)$ NLSM, which also ensures that the $U(N)$ NLSM considered in Ref. \cite{Cachazo:2016njl} has the same amplitudes as $SU(N)$ NLSM.

\subsection{NLSM of coset $H\times H/H$}
\label{app:diagnlsm}
For the symmetric coset $H_L \times H_R / H_V$ where $H_L = H_R = H_V = H$, the coset space is isomorphic to the space of group $H$. Such a property can be used to construct the NLSM Lagrangian. An element in the unbroken diagonal group $H_V$ can be parameterized using $U \equiv g_R g_L^{-1}$ \cite{Kampf:2013vha}, where $g_{L/R}$ are elements of $H_{L/R}$, and $g_L = g_R$. Then the global transformation of $U$ under the broken group $H_L \times H_R$ is given by
\bea
U \to V_R U V_L^{-1},
\eea
where $V_{L/R}$ are transformation matrices of $H_{L/R}$. As the transformation is linear, the two derivative Lagrangian can simply be written as
\bea
\lag_{\nlsm}^{(2)} = \frac{f^2}{8} \tr \left\{ \partial_\mu U \partial^\mu U^{-1} \right\}.
\eea
We need to write $U$ using the exponential map of $H_V$, which is parameterized by the NGBs:
\bea
U = \exp \left( i \frac{2 \pi^a T^a}{f} \right),\label{eq:a3gud}
\eea
where $(T^a)_{bc} = -if^{abc}$ are the generators of $H_V$. We see that the NGBs $\pi^a$ are in the adjoint representation of $H_V$.

\subsection{Parameterization of $U(N)$ NLSM}
\label{app:nlsmpara}
In the NLSM Lagrangian, in general we can redefine $\pi \to \pi  + \ordr (\pi^2)$, which will change the form of the Lagrangian and the Feynman rules, but not the amplitudes. In other words, we can have different ways to parameterize the NGBs, which does not change the physics. The parameterization in Eq. (\ref{eq:a3gud}) is commonly referred to as the ``exponential parameterization''; the infrared construction of NLSM discussed in Section \ref{sec:nlsm} also leads to such a parameterization. It was noticed early on that for $H_V = U(N)$, a group of general parameterizations is given by \cite{Cronin:1967jq}
\bea
U \left( i \frac{2 \pi^a T^a}{f} \right) =1+ \sum_{k=1}^{\infty} a_k \left( i \frac{2 \pi^a T^a}{f} \right)^k,
\eea
where the coefficients $a_k$ are arbitrary as long as the resulting $U$ is unitary. Namely, the function $U$ satisfies $[U (\underline{x})]^{-1} = U(-\underline{x})$, and the constraint on the coefficients $a_k$ is
\bea
\sum_{k=0}^n a_k a_{n-k} (-1)^k = 0
\eea
for $n>0$, with $a_0 \equiv 1$. The above constraint is non-trivial for an even $n$. The exponential parameterization corresponds to $a_k = 1/k!$.

The Cayley parameterization, which is also often used, is given by the function
\bea
U(\underline{x}) = \frac{1+(\underline{x}/2)}{1-(\underline{x}/2)},
\eea
so that $a_k = 2^{1-k}$. Such a parameterization leads to the very simple flavor-ordered Feynman rules of NLSM:
\bea
V^{\nlsm}_{2k+1}(\mathbb{I}_{2k+1}) = 0, \qquad V^{\nlsm}_{2k+2} (\mathbb{I}_{2k+2}) = \frac{i(-1)^k}{f^{2k}} \left( \sum_{i=0}^k p_{2i+1} \right)^2\ ,
\label{eq:cayleyvec}
\eea
which enables us to derive the subleading single and triple soft theorems of NLSM by directly evaluating the Feynman diagrams \cite{Low:2018acv}. In the extended theory $\nlsm \oplus \phi^3$ under Cayley parameterization, the odd-pt vertices with three $\phi$ is given by
\bea
V^{\text{\nlsm} \oplus \phi^3} (\mathbb{I}_{2k+1} | 1,2k+1,j)  = \left\{ \begin{array}{ll}
\frac{i(-1)^k}{f^{2k}} & \text{for even } j,\\
0& \text{for odd }j.
\end{array} \right.\label{eqnlsmclfr}
\eea

\section{Review of CHY formulae for scalar EFTs}
\label{app:chyr}
The Cachazo-He-Yuan formalism can be applied to write down the tree-level amplitudes of a variety of single-parameter theories, including the scalar EFTs discussed in this paper. In general, the amplitude for a scalar theory is written as
\bea
M_n  = \oint d\mu_n \; \mathcal{I}_L (\{p, \sigma \}) \; \mathcal{I}_R (\{p, \sigma\}),
\eea
where $p$ denotes the on-shell momenta, and the integral is on the dimensionless variables $\sigma_i$ satisfying the scattering equation
\bea
E_a \equiv \sum_{i \ne a} \frac{p_i \cdot p_a}{ \sigma_{ia} }=0,
\eea
with $\sigma_{ij} \equiv \sigma_i - \sigma_j$. The measure $d \mu_n$, defined as
\bea
d\mu_n &\equiv& ( \sigma_{ij}\sigma_{jk}\sigma_{ki}) (\sigma_{pq}\sigma_{qr}\sigma_{rp}) \prod_{a \neq i,j,k} E_a^{-1} \prod_{b \neq p,q,r} d\sigma_b,
\eea
contains $2(3-n)$ powers of momenta. The integrands $\mathcal{I}_L$ and $\mathcal{I}_R$ vary among different theories.

The theories we need to consider are the biadjoint $\phi^3$ theory, NLSM, special Galileon and the extensions of these theories. The building blocks include the Parke-Taylor factor
\bea
\mathcal{C}_n (\omega) = \frac{1}{\sigma_{\omega_1 \omega_2} \cdots \sigma_{\omega_{n-1} \omega_n} \sigma_{\omega_n \omega_1}},
\eea
where $\omega$ is the ordering of the flavor indices; and the Pfaffian of the anti-symmetric matrix
\bea
[\A_n]_{ab} =\left\{\begin{array}{ll} 	\dfrac{2 k_a \cdot k_b}{\sigma_{ab}}, & a \neq b, \\
	0, & a = b.\end{array} \right.
	\eea
$\pf \A_n$ contains $n$ powers of momenta.

The most general single trace amplitude written using these ingredients is
\bea
M^{\pi+}_n (\alpha | \beta)= \oint d\mu_n \; \Big( \mathcal{C}(\alpha)\, (\pf\, \A_{\bar{\alpha}})^2 \Big) \Big( \mathcal{C}(\beta)\, (\pf\, \A_{\bar{\beta}})^2 \Big),\label{eq:chyeg}
\eea
which is the extended theory $\pi+$ where particles of labels within $\alpha \cap \beta$ are biadjoint scalars, those in $\alpha \cap \bar{\beta}$ or $\bar{\alpha} \cap \beta$ are NGBs $\gs$ or $\gst$,  and those in $\bar{\alpha} \cap \bar{\beta}$ are special Galileons.
The power of momenta in such an amplitude is given by $m=2(3+n - \nf_\alpha - \nf_\beta)$, where $\nf_{\alpha /\beta}$ are the number of labels in $\alpha/\beta$.

It is understood that $(\pf\, \A_{\bar{\alpha}}) \equiv 1$ when $\bar{\alpha} = \varnothing$, or $\nf_\alpha = n$. When $\nf_\alpha < n$ but $\nf_{\beta}  = n$, Eq. (\ref{eq:chyeg}) is reduced to the amplitude of the extended theory of NLSM:
\bea
M^{\nlsm \oplus \phi^3}_n (\alpha | \beta) =\oint d\mu_n \;  \Big( \mathcal{C}(\alpha)\, (\pf\, \A_{\bar{\alpha}})^2 \Big) \mathcal{C}(\beta).
\eea
Here we see that $\nlsm \oplus \phi^3$ is naturally part of the full theory $\pi+$. When $\nf_{\alpha} = \nf_{\beta} =  n$, Eq. (\ref{eq:chyeg}) becomes an amplitude of pure biadjoint scalars:
\bea
M^{\phi^3}_n (\alpha | \beta)= \oint d\mu_n \;  \mathcal{C}(\alpha)  \mathcal{C}(\beta).
\eea

If we remove the ordering $\beta$, Eq. (\ref{eq:chyeg}) needs to be modified to
\bea
M^{\text{sGal} \oplus \nlsm}_n (\alpha) =  \oint d\mu_n~\Big( \mathcal{C}_n  (\alpha)(\pf \A_{\bar{\alpha}})^2 \Big)~(\pf' \A_n)^2\label{eq:chysgg}
\eea
which is an amplitude with special Galileons interacting with NGBs. The reduced Pfaffian is defined as $\pf' \A_n = \frac{(-)^{a+b}}{\sigma_{ab}} \pf \A_n^{[a,b]}$, where $\A_n^{[a,b]}$ is the matrix $\A_n$ with rows and columns of labels $a$ and $b$ removed. Such an amplitude contains $m = 2(1 + n -\nf_\alpha)$ powers of momenta. Again, when $\nf_\alpha = n$, Eq. (\ref{eq:chysgg}) is reduced to pure NLSM amplitude
\bea
M^{\nlsm }_n (\alpha) =  \oint d\mu_n~\mathcal{C}_n  (\alpha) ~(\pf' \A_n)^2.
\eea
The last thing we can construct is an amplitude with no flavor labels, which is the pure special Galileon amplitude
\bea
M^{\sgal }_n =  \oint d\mu_n~(\pf' \A_n)^4
\eea
with $m=2(n-1)$ powers of momenta.

The flavor-ordered amplitudes presented above are all single-trace ones. Recently Ref. \cite{Mizera:2018jbh} proposed the CHY formulae for multi-trace amplitudes in $\nlsm \oplus \phi^3$. Those amplitudes do not appear in the subleading single soft theorem of NLSM, but they certainly offer more consistency checks for the Lagrangian we write down in Section \ref{sec:elag}, which we leave for future work. It should also be understood that the above representation only gives the kinematic and flavor structure of an amplitude; it does not contain information like the coupling strength, thus may not give an amplitude of correct mass dimension in our convention of only stripping the dimensionless flavor structure when writing down flavor-ordered amplitudes. An amplitude of $n$ external scalars should have a mass dimension of $4-n$, which in general does not agree with the $m$ values presented above.

\section{Identification of Feynman vertices for the extension of special Galileon}

\label{app:sgaled}

\begin{figure}[htbp]
\centering
\includegraphics[width=0.42\textwidth]{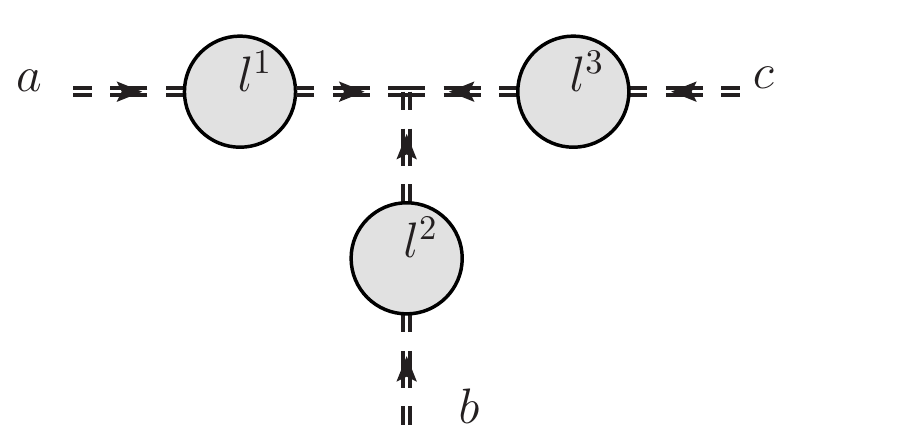}
\qquad
\includegraphics[width=0.33\textwidth]{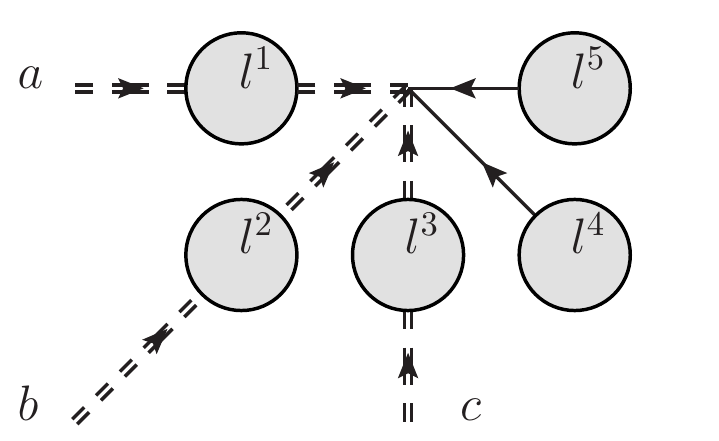}
\caption{ The two kinds of diagrams in $M_{n-1}^{\pi+} (a,b,c|a,b,c)$. We use double dashed lines to represent $\phi$, and solid lines to represent the special Galileon $\pi$. The blobs represent semi-on-shell sub-amplitudes. We omit the external $\pi$ lines.
}\label{figsgalabc}
\end{figure}

\begin{figure}[htbp]
\centering
\includegraphics[width=0.42\textwidth]{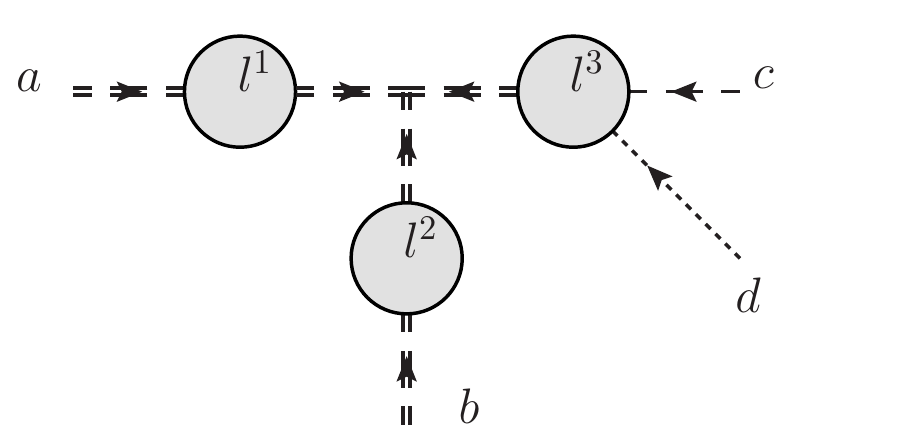}
\qquad
\includegraphics[width=0.33\textwidth]{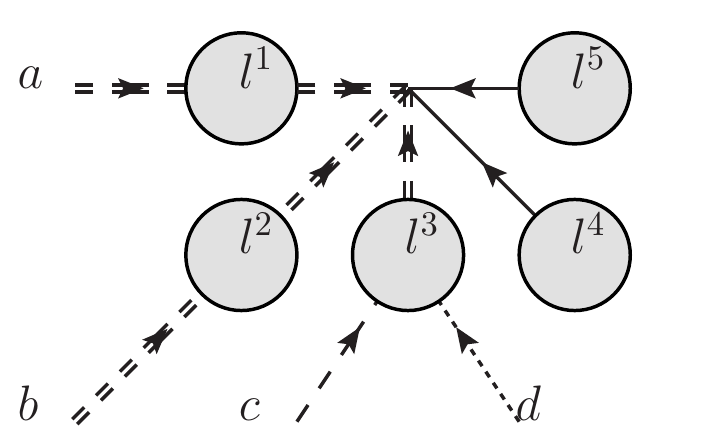}
\qquad
\includegraphics[width=0.42\textwidth]{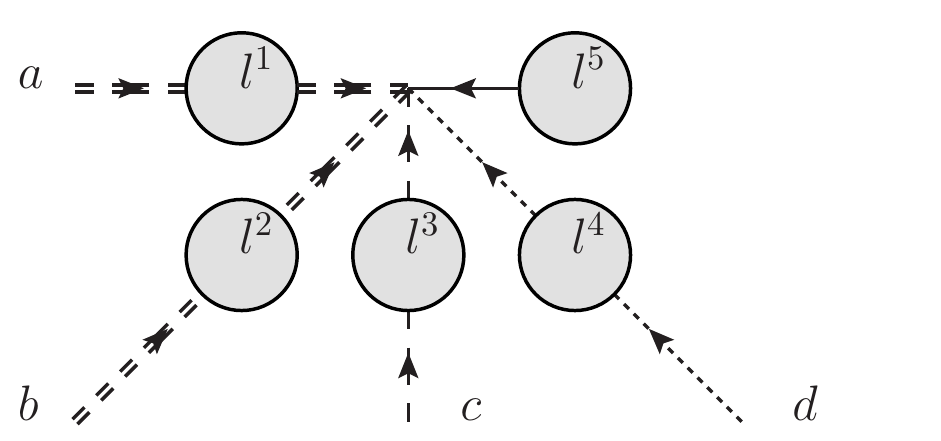}
\caption{ The three kinds of diagrams in $M_{n-1}^{\pi+} (a,b,c|a,b,d)$. We use single dashed lines to represent $\Sigma$, and dotted lines to represent $\tilde{\Sigma}$.
}\label{figsgalad}
\end{figure}

\begin{figure}[htbp]
\centering
\includegraphics[width=0.42\textwidth]{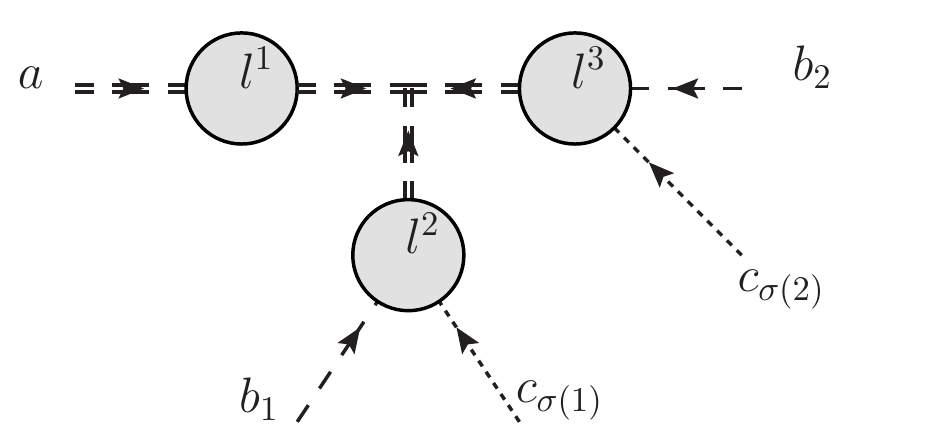}
\qquad
\includegraphics[width=0.33\textwidth]{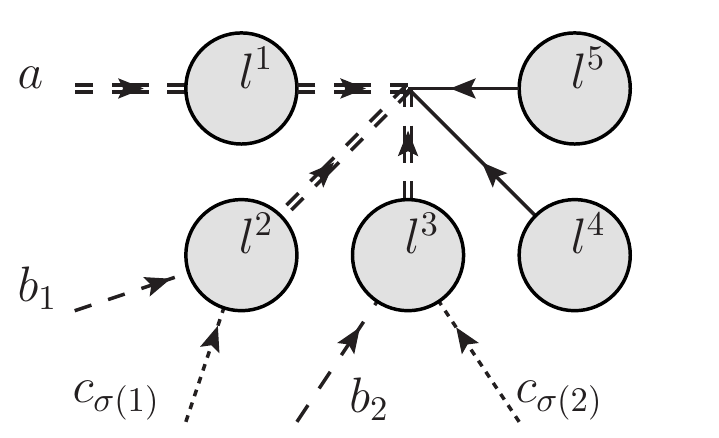}
\qquad
\includegraphics[width=0.42\textwidth]{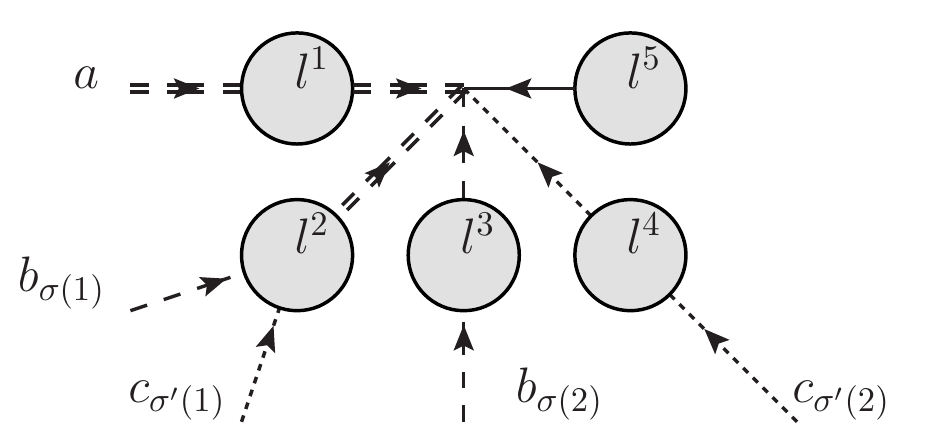}
\qquad
\includegraphics[width=0.42\textwidth]{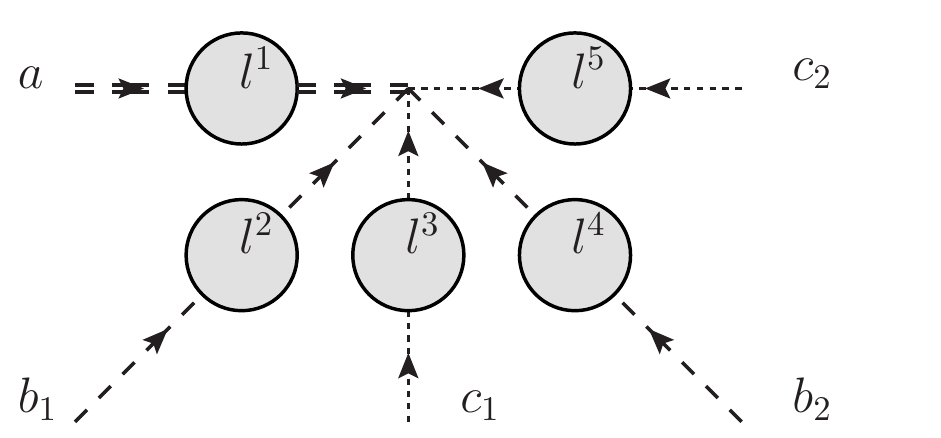}
\caption{ The four kinds of diagrams in $M_{n-1}^{\pi+} (a,b_1,c_1|a,b_2,c_2)$.
}\label{figsgalae}
\end{figure}

The Ward identity corresponding to the enhanced shift symmetry of special Galileon leads to the soft theorem given by Eq. (\ref{eq:sgaltsfw}), on the RHS of which we see products of three or five semi-on-shell sub-amplitudes. Below we will use the following shorthand notation to show these terms:
\bea
J^m (l) \equiv \prod_{i=1}^m J ( \{ p_{l^i_j} \}),
\eea
where $m = 3,5$.
The soft theorem given by the CHY formalism, on the other hand, is Eq. (\ref{eqsgalchy}), on the RHS of which we have amplitudes of three different kinds of particle content. We now write these amplitudes using the conjectures given in Section \ref{sec:sgale}. The first kind of amplitude has 3 $\phi$'s and $n-4$ Galileons:
\bea
iM_{n-1}^{\pi+} (a,b,c|a,b,c) &=& \sum_{l: C^3(a,b,c)} V^{\pi+}_3(q_{l^1},q_{l^2},q_{l^3}|q_{l^1},q_{l^2},q_{l^3}) J^3 (l)\non\\
&&+\sum_{l: C^3(a,b,c)} V^{\pi+}_5(q_{l^1},q_{l^2},q_{l^3}|q_{l^1},q_{l^2},q_{l^3}) J^5 (l),\label{eqsgalmabc}
\eea
where the sum over $l:C$ means summing over ways to devide $\{1,2, \cdots, n-1\}$ into $m$ disjoint, non-ordered subsets that satisfies condition $C$, and $C^3(a,b,c) = \{a \in l^1, b \in l^2, c \in l^3\}$. The two terms in Eq. (\ref{eqsgalmabc}) corresponds to the two kinds of diagrams shown in Fig. \ref{figsgalabc}. The second kind of amplitude has two $\phi$'s, one $\Sigma$, one $\tilde{\Sigma}$ and $n-5$ Galileons:
\bea
iM_{n-1}^{\pi+} (a,b,c|a,b,d) &=& \sum_{l:C^4_1 (a,b,c,d)} V^{\pi+}_3(q_{l^1},q_{l^2},q_{l^3}|q_{l^1},q_{l^2},q_{l^3}) J^3 (l)\non\\
&&+\sum_{l: C^4_1 (a,b,c,d)} V^{\pi+}_5(q_{l^1},q_{l^2},q_{l^3}|q_{l^1},q_{l^2},q_{l^3}) J^5 (l)\non\\
&&+ \sum_{l: C^4_2 (a,b,c,d)} V^{\pi+}_5(q_{l^1},q_{l^2},q_{l^3}|q_{l^1},q_{l^2},q_{l^4}) J^5 (l),
\eea
where $C^4_1 (a,b,c,d) = \{ a \in l^1,b \in l^2, \{c,d\} \subset l^3 \}$, $C^4_2 (a,b,c,d) =\{ a\in l^1, b\in l^2, c\in l^3,d \in l^4\}$, and the three terms come from the three diagrams shown in Fig. \ref{figsgalad}. The third kind of amplitude has one $\phi$, two $\Sigma$'s, two $\tilde{\Sigma}$'s, and $n-6$ Galileons:
\bea
iM_{n-1}^{\pi+} (a,b_1,b_2|a,c_1,c_2) &=& \sum_{\sigma \in S_2} \sum_{l: C^5_1(a,b_1,c_{\sigma(1)},b_2,c_{\sigma(2)})} V^{\pi+}_3(q_{l^1},q_{l^2},q_{l^3}|q_{l^1},q_{l^2},q_{l^3}) J^3 (l)\non\\
&&+\sum_{\sigma \in S_2} \sum_{l: C^5_1(a,b_1,c_{\sigma(1)},b_2,c_{\sigma(2)})} V^{\pi+}_5(q_{l^1},q_{l^2},q_{l^3}|q_{l^1},q_{l^2},q_{l^3}) J^5 (l)\non\\
&&+\sum_{\sigma,\sigma' \in S_2} \sum_{ l:C^5_2(a,b_{\sigma(1)} ,c_{\sigma'(1)}, b_{\sigma(2)}, c_{\sigma'(2)})}V^{\pi+}_5(q_{l^1},q_{l^2},q_{l^3}|q_{l^1},q_{l^2},q_{l^4}) J^5 (l)\non\\
&&+ \sum_{l:C^5_3 (a,b_1, c_1, b_2, c_2)}V^{\pi+}_5(q_{l^1},q_{l^2},q_{l^3}|q_{l^1},q_{l^4},q_{l^5}) J^5 (l) ,
\eea
where $C^5_1(a,b,c,d,e) =\{ a\in l^1, \{b, c \} \subset l^2,  \{d,e \} \subset l^3 \}$, $C^5_2(a,b,c,d,e) =\{ a \in l_1, \{b, c\} \subset l^2, d \in l^3, e \in l^4\}$, $C^5_3 (a,b,c,d,e) = \{a \in l^1, b \in l^2, c \in l^3, d \in l^4, e \in l^5 \}$, and the four terms come from the four diagrams shown in Fig. \ref{figsgalae}.
%
%
%
%
%
%
%
%
%
%
%
%

We now express the RHS of Eq. (\ref{eqsgalchy}) using the three kinds of amplitudes discussed above:
\bea
&&\sum_{a=2}^{n-2} \sum_{\substack{c=2 \\ c \neq a}}^{n-1}  \sum_{\substack{d=1 \\ d \neq a}}^{n-2} s_{an} s_{cn} s_{dn} M_{n-1}^{\pi +} (a ,c ,1| n-1, d, a) \non\\
&=& \sum_{1<i_1<i_2<i_3<n-1} s_{i_1 n} s_{i_2 n} s_{i_3 n} \sum_{\sigma \in S_{3}} M_{n-1}^{\pi +} (i_{\sigma (1)} ,i_{\sigma (2)},1| n-1, i_{\sigma (3)}, i_{\sigma (1)})\non \\
&&+ \sum_{1<i_1<i_2<n-1} s_{i_1 n} s_{i_2 n} s_{1 n} \sum_{\sigma \in S_{2}} M_{n-1}^{\pi +} (i_{\sigma (1)} ,i_{\sigma (2)},1| n-1,1, i_{\sigma (1)})\non\\
&&+\sum_{1<i_1<i_2<n-1} s_{i_1 n} s_{i_2 n} s_{(n-1) n} \sum_{\sigma \in S_{2}} M_{n-1}^{\pi +} (i_{\sigma (1)} ,n-1,1| n-1, i_{\sigma (2)}, i_{\sigma (1)})\non\\
&&+\sum_{1<i<n-1} s_{in} s_{1n} s_{(n-1)n} M^{\pi+}_{n-1} (i,n-1,1|n-1,1,i)\non\\
&&+\sum_{1<i_1<i_2<n-1} \sum_{\sigma \in S_{2}} s_{i_{\sigma(1)} n} s^2_{i_{\sigma(2)} n} A_{n-1}^{\pi +} (i_{\sigma (1)} ,i_{\sigma (2)},1| n-1, i_{\sigma (2)}, i_{\sigma (1)}).
\eea
We also have the useful relation
\bea
&&\sum_{1<i_1<i_2<n-1} \sum_{\sigma \in S_{2}} s_{i_{\sigma(1)} n} s^2_{i_{\sigma(2)} n} M_{n-1}^{\pi +} (i_{\sigma (1)} ,i_{\sigma (2)},1| n-1, i_{\sigma (2)}, i_{\sigma (1)})\non\\
&=& \sum_{1<i_1<i_2<n-1} s_{i_1 n} s_{i_2 n} (s_{i_1 n} + s_{i_2 n}) M_{n-1}^{\pi +} (i_{1} ,i_2,1| n-1, i_2, i_1)\non\\
&=&\sum_{1<i_1<i_2<i_3<n-1} \frac{1}{2}s_{i_1 n} s_{i_2 n} s_{i_3 n} \sum_{\sigma \in S_{3}} M_{n-1}^{\pi +} (i_{\sigma (1)} ,i_{\sigma (2)},1| n-1, i_{\sigma (1)}, i_{\sigma (2)}) \non\\
&&+ \sum_{1<i_1<i_2<n-1}s_{i_1 n} s_{i_2 n} (s_{1n} + s_{(n-1)n}) M_{n-1}^{\pi +} (i_{1} ,i_2,1| n-1, i_1, i_2) ,
\eea
where we use total momentum conservation as well as the on-shell condition of the external momenta. Putting everything together, we arrive at
\bea
i\sum_{a=2}^{n-2} \sum_{\substack{c=2 \\ c \neq a}}^{n-1}  \sum_{\substack{d=1 \\ d \neq a}}^{n-2} s_{an} s_{cn} s_{dn} M_{n-1}^{\pi +} (a ,c ,1| n-1, d, a) = \sum_l G_3 (l) J^3(l) + \sum_l G_5 (l) J^5(l),
\eea
where
\bea
G_3(l) = 8 p_{q,l^1} p_{q,l^2} p_{q,l^3} V_3^{\pi+} (q_{l^1}, q_{l^2}, q_{l^3}|q_{l^1}, q_{l^2}, q_{l^3}),
\eea
and
\bea
\sum_l G_5(l) J^5(l) &=&8 \left\{ \sum_{l:\{1,n-1\} \subset l^5} \sum_{\sigma \in S_4}\left[ p_{q,l^{\sigma(1)}} p_{q,l^{\sigma(2)}} p_{q,l^{\sigma(3)}} V_5^{(1)} (q_{l^{\sigma(1)}}, q_{l^{\sigma(2)}}, q_{l^{\sigma(3)}},q_{l^{\sigma(4)}},q_{l^5})\right.\right.\non\\
&&\left.+\frac{1}{4}p_{q,l^{\sigma(1)}} p_{q,l^{\sigma(2)}} p_{q,l^5} V_5^{\pi+} (q_{l^{\sigma(1)}},q_{ l^{\sigma(2)}}, q_{l^5}|q_{l^{\sigma(1)}},q_{ l^{\sigma(2)}}, q_{l^5})\right]\non\\
&&+\sum_{l:1\in l^4, n-1 \in l^5} \sum_{\sigma \in S_3} \left[ p_{q,l^{\sigma(1)}} p_{q,l^{\sigma(2)}} p_{q,l^{\sigma(3)}} V_5^{(2)} (q_{l^{\sigma(1)}}, q_{l^{\sigma(2)}}, q_{l^{\sigma(3)}},q_{l^4}, q_{l^5})\right.\non\\
&&+ p_{q,l^{\sigma(1)}} p_{q,l^{\sigma(2)}} p_{q,l^{4}} V_5^{(3)} (q_{l^{\sigma(1)}}, q_{l^{\sigma(2)}}, q_{l^{\sigma(3)}},q_{l^4}, q_{l^5})\non\\
&&+ p_{q,l^{\sigma(1)}} p_{q,l^{\sigma(2)}} p_{q,l^{5}} V_5^{(4)} (q_{l^{\sigma(1)}}, q_{l^{\sigma(2)}}, q_{l^{\sigma(3)}},q_{l^4}, q_{l^5}) \non\\
 && \left.  \left.+\frac{1}{2}p_{q,l^{\sigma(1)}} p_{q,l^4} p_{q,l^{5}} V_5^{\pi+} (q_{l^{\sigma(1)}},q_{ l^{4}}, q_{l^5}|q_{l^{\sigma(1)}},q_{ l^{4}}, q_{l^5}) \right] J^5 (l)\right\} ,
\eea
with
\bea
V_5^{(1)} (1, 2,3,4,5) &=& \frac{1}{2} V_5^{\pi^+} (1, 2,5|1, 2,5)   - V_5^{\pi^+} (1, 2, 5|1, 3, 5) ,\\
V_5^{(2)} (1,2,3,4,5) &=& \frac{1}{2}  V_5^{\pi^+} (1, 2, 4| 1, 2, 5 ) - V_5^{\pi^+} (1,2, 4|1, 3, 5 ),\\
V_5^{(3)}  (1, 2,3,4,5) &=& \frac{1}{2} V_5^{\pi^+} (1, 2,4|1, 2,5)   + V_5^{\pi^+} (1, 2, 4|1, 5,4) ,\\
V_5^{(4)}  (1, 2,3,4,5) &=& \frac{1}{2} V_5^{\pi^+} (1, 2,4|1, 2,5)   + V_5^{\pi^+} (1, 4,5|1, 2,5) .
\eea

Our result Eq. (\ref{eq:sgaltsfw}) derived from the Ward identity gives us the following constraints:
\bea
G_3 (l ) &=&  8i\lambda  p_{q, l^{1}} p_{q, l^{2}} p_{q, l^{3}},
\eea
and
\bea
\sum_l G_5(l) J^5(l) = - i \frac{2\lambda}{3 \alpha^2} \sum_{l}   \sum_{\sigma \in S_5}   p_{q, l^{\sigma(1)}} p_{q, l^{\sigma(2)}} p_{q, l^{\sigma(3)}} \left[q_{l^{\sigma(4)}}^2 q_{l^{\sigma(5)}}^2 - p_{l^{\sigma(4)}, l^{\sigma(5)}}^2 \right] J^5 (l),
\eea
thus the constraints on the vertices are
\bea
i \lambda  &=& V_3^{\pi+} (1,2,3|1,2,3),\\
 -\frac{i\lambda}{ \alpha^2} \left[ p_4^2 p_5^2 - p_{4,5}^2 \right]&=& V_5^{\pi+} (1,2,3|1,2,3) = \sum_{\sigma \in S_3} V_5^{(1)} ( \sigma (1), \sigma (2), \sigma (3 ),4,5) \non\\
 &=& \sum_{\sigma \in S_3} V_5^{(2)} ( \sigma (1), \sigma (2), \sigma (3 ),4,5)\non\\
 &=& \sum_{\sigma \in S_2} V_5^{(3)} ( \sigma (1), \sigma (2) ,4,\sigma (3),5)\non\\
 &=& \sum_{\sigma \in S_2} V_5^{(4)} ( \sigma (1), \sigma (2) ,5,\sigma (3),4).\label{eqsgalocv5}
\eea
Now we can directly get $V_3^{\pi+} (1,2,3|1,2,3)$ and $V_5^{\pi+} (1,2,3|1,2,3)$, and Eq. (\ref{eqsgalocv5}) also imposes constraints on $V_5^{\pi+} (1,2,3|1,2,4)$ and $V_5^{\pi+} (1,2,3|1,4,5)$. The constraints are rather weak:
\bea
\sum_{\sigma \in S_2} V_5^{(3)} ( \sigma (1), \sigma (2) ,4,\sigma (3),5) = \sum_{\sigma \in S_2} V_5^{(4)} ( \sigma (1), \sigma (2) ,5,\sigma (3),4)
\eea
is automatically satisfied because of the symmetry of the left and right indices, and
\bea
\sum_{\sigma \in S_3} V_5^{(1)} ( \sigma (1), \sigma (2), \sigma (3 ),4,5)  = \sum_{\sigma \in S_2} V_5^{(3)} ( \sigma (1), \sigma (2) ,4,\sigma (3),5)
\eea
is not an independent constraint either, so that the only useful constraints are
\bea
\sum_{\sigma \in S_3} V_5^{(2)} ( \sigma (1), \sigma (2), \sigma (3 ),4,5) &=&-\frac{i\lambda}{ \alpha^2} \left[ p_4^2 p_5^2 - p_{4,5}^2 \right],\label{eqsgalfcae}\\
\sum_{\sigma \in S_2} V_5^{(3)} ( \sigma (1), \sigma (2) ,4,\sigma (3),5) &=&-\frac{i\lambda}{ \alpha^2} \left[ p_4^2 p_5^2 - p_{4,5}^2 \right].\label{eqsgalfcad}
\eea
Then we have some freedom to choose the specific form of the vertices. The two set of solutions presented in Section \ref{sec:sgale} are examples satisfying the above constraints.
\end{appendix}

\bibliographystyle{JHEP}
\bibliography{references_amp}

\end{document}